\documentclass[twocolumn,floatfix]{revtex4}
\usepackage[utf8]{inputenc}
\usepackage{graphicx}%
\usepackage{amsmath}%
\usepackage{amsfonts}%
\usepackage{amssymb}
\usepackage{bm}
\usepackage{color}

\def\d{{\partial}}
\def\s{{\sigma}}
\def\e{{\epsilon}}
\def\k{{ {\bm k} }}
\def\p{{ {\bm p} }}
\def\q{{ {\bm q} }}
\def\Q{{ {\bm Q} }}
\def\0{{ {\bm 0} }}

\def\w{{\omega}}
\def\a{{\alpha}}
\def\b{{\beta}}

\def\g{{\gamma}}
\def\l{{\lambda}}
\def\r{{ {\bm r} }}

\allowdisplaybreaks[4]

\begin{document}
\title{
Odd-parity spin-loop-current order mediated by transverse spin fluctuations \\
in cuprates and related electron systems
}
\author{
Hiroshi Kontani,
Youichi Yamakawa,
Rina Tazai, and
Seiichiro Onari 
}

\date{\today}

\begin{abstract}
Unconventional symmetry-breaking phenomena due to 
nontrivial order parameters attract increasing attention in 
strongly correlated electron systems.
Here, we predict theoretically the occurrence of nanoscale spontaneous
spin-current, called the spin loop-current (sLC) order,
as a promising origin of the pseudogap and electronic nematicity in cuprates. 
We reveal that the sLC is driven by the
odd-parity electron-hole condensation that are mediated 
by transverse spin fluctuations around the pseudogap temperature $T^*$.
At the same temperature, 
odd-parity magnon pair condensation occurs.
The sLC order is ``hidden'' in that neither internal magnetic field nor 
charge density modulation is induced, whereas the predicted sLC 
with finite wavenumber 
naturally gives the Fermi arc structure. 
In addition, the fluctuations of sLC order work as attractive 
pairing interaction between adjacent hot spots, which enlarges 
the $d$-wave superconducting transition temperature $T_{\rm c}$.
The sLC state will be a key ingredient in understanding the pseudogap,
electronic nematicity as well as superconductivity
in cuprates and other strongly correlated metals.

\end{abstract}

\address{
Department of Physics, Nagoya 
Furo-cho, Nagoya 464-8602, Japan. 
}
 
\sloppy

\maketitle

%%%%%%%%%%%%%%%%%%
%Introduction
\section{Introduction}
%%%%%%%%%%%%%%%%%%
%various symmetry breaking
% TRS breaking
% \section{Introduction}
%Recently, very rich symmetry breaking phenomena
%have been discovered in many strongly correlated metals,
%and their microscopic mechanisms have been studied very actively.

\begin{figure}[htb]
%\vspace{5mm}
\includegraphics[width=.90\linewidth]{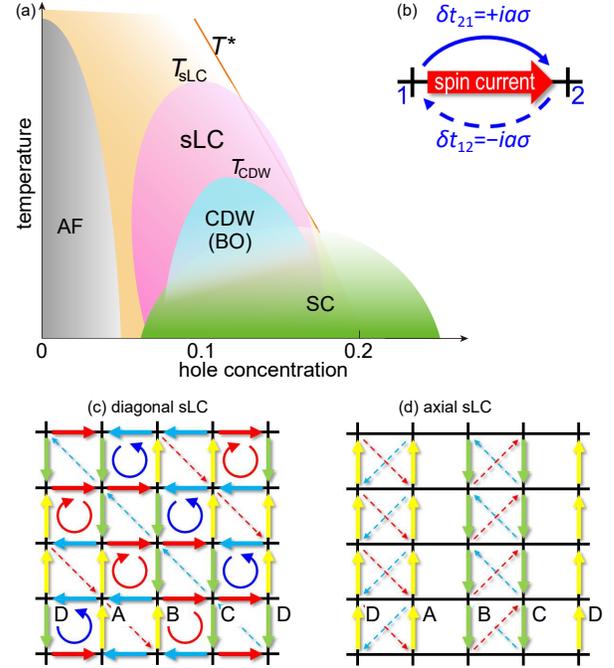}
\caption{
(a) Possible phase-diagram of 
hole-doped cuprate superconductors.
The sLC phase is obtained by the present study.
(b) Current of spin ($\s=\pm1$) from site 1 to site 2.
$\s=\pm1$ is the spin of the electron.
(c)(d) Schematic pictures of the
diagonal sLC at $\q_{\rm sLC}=(\pi/2,\pi/2)$
and the axial sLC at $\q_{\rm sLC}=(\pi/2,0)$, respectively.
}
\label{fig:fig1}
\end{figure}

Various unconventional symmetry-breaking phenomena,
such as violations of rotational and parity symmetries,
have been discovered in many strongly correlated metals recently.
This fact strongly indicates the 
emergence of exotic density-wave orders,
which are totally different from usual spin/charge density waves. 
Various exotic symmetry-breaking phenomena,
such as violations of rotational and parity symmetries,
are the central issues in cuprate high-$T_{\rm c}$ superconductors.
However, their microscopic mechanisms still remain as unsolved issues.
Figure \ref{fig:fig1} (a) shows a schematic phase diagram of 
cuprate superconductors.
Below $T_{\rm CDW}\sim 200$K,
a stripe charge-channel density-wave emerges 
at finite wavevector $\q\approx(\pi/2,0)$ in many compounds
\cite{Y-Xray1,Bi-Xray1,STM-Kohsaka,STM-Fujita},
which produces the Fermi arc structure and 
causes a reduction in the density-of-states (DOS).
However, it cannot be the origin of the pseudogap temperature $T^*$
since $T^*>T_{\rm CDW}$. 
Short quasiparticle lifetime due to spin or charge fluctuations 
could reduce the DOS
% around the Fermi level
\cite{Tremblay,Scalapino,Moriya}.

%%%%%%%%%%%%%%%%%%%%
%%%%%%%%%%%%%%%%%%%

Recently, much experimental evidence for the phase transition at
$T^*$ has been accumulated 
\cite{RUS,ARPES-Science2011,Y-Sato,Hg-Murayama,Fujimori-nematic,Shibauchi-nematic}.
%The origin of density-wave and nematic order
Various fascinating order parameters
have been proposed and actively investigated,
such as the CDW or bond-order (BO)
\cite{Bulut,Chubukov,Chubukov-AL,Sachdev,Metzner,DHLee-PNAS,Kivelson-NJP,Yamakawa-CDW,Tsuchiizu-CDW,Kawaguchi-CDW}, 
the pair-density-wave
\cite{PALee,Agterberg},
and the charge loop-current (cLC) order
\cite{Varma,Affleck,FCZhang,Schultz}.
%\cite{Varma,Affleck,Schultz,FCZhang,Bulut-cLC,Yokoyama}.
%However, the true order parameter(s) are still unknown.
%These exotic orders cannot be described
%with a standard Hubbard model with on-site Coulomb interactions $U$
%using mean-field-level approximations.
%add
At present, the symmetries of the hidden order
in the pseudo-gap phase are not yet confirmed experimentally.
%understood only partially.
Thus, it is necessary to study various possibilities without prejudice
based on advanced many-body theories
%Therefore, various possibiliries should be 
%To understand the origin of these exotic orders,
%higher-order quantum processes have been considered actively
\cite{Chubukov,Chubukov-AL,Sachdev,Metzner,DHLee-PNAS,Kivelson-NJP,Yamakawa-CDW,Tsuchiizu-CDW,Kawaguchi-CDW,PALee,Agterberg,Varma,Affleck,FCZhang,Schultz}.
%since simple mean-field-level approximations
%usually lead to conventional spin-density-wave instabilities.

Let us discuss the symmetry breaking in the correlated hopping
between sites $i$ and $j$;
$t_{i,j} \rightarrow t_{i,j}+\delta t_{i,j}$,
where $\delta t_{i,j} (=\delta t_{j,i})^*$ is the order parameter.
Then, the BO is given by a real and even-parity $\delta t_{i,j}$
\cite{Bulut,Chubukov,Sachdev,Metzner,DHLee-PNAS,Kivelson-NJP,Yamakawa-CDW,Tsuchiizu-CDW,Kawaguchi-CDW}.
A spin-fluctuation mechanism
\cite{Tsuchiizu-CDW,Kawaguchi-CDW}
predicts the ferro ($\q={\bm 0}$) $d$-wave BO state at $T^*$
and stripe ($\q\approx(\pi/2,0)$) BO at $T_{\rm CDW}$.
The former order explains the experimental nematic transition
\cite{Y-Sato,Shibauchi-nematic}.
However, simple translational symmetry preserving ferro-BO 
does not explain the pseudogap formation.
% https://www.pnas.org/content/116/27/13249?ijkey=d4aebf42f56e44c5c0579b6caf696a3eaaecb6d2&keytype2=tf_ipsecsha
O% M. H. Hamidian et al., Detection of a Cooper-pair density wave in Bi2Sr2CaCu2O8+x. Nature 532, 343 (2016)
Also, the cLC order is given by a pure imaginary and odd-parity 
$\delta t_{i,j}$
\cite{Varma,Affleck,FCZhang,Schultz}.
Both order parameters have been actively investigated.

In contrast, spin current flows if pure imaginary order parameter
is odd under space and spin inversions;
$\delta t_{i,j}^{\s}=-\delta t_{j,i}^{\s}=-\delta t_{i,j}^{-\s}$
as shown in Fig. \ref{fig:fig1} (b)
\cite{Schultz,Nersesyan,Ozaki,Ikeda,Fujimoto,Sr2IrO4}.
Here, $\s=\pm 1$ represents the spin of the electron.
Figures \ref{fig:fig1} (c) and (d) depict 
the spin loop-current (sLC) order
at the wavevectors $\q_{\rm sLC}=(\delta,\delta)$
and that at $\q_{\rm sLC}=(\delta,0)$
with $\delta=\pi/2$, respectively.
The sLC is a hidden order in the sense that
no internal magnetic field appears,
and charge density modulation is quite small.
Nonetheless, the sLC is very attracting since 
the pseudogap and Fermi surface (FS) reconstruction
are induced by band-folding if $\q_{\rm sLC}\ne{\bm0}$.

In this paper,
we discover the emergence of ``hidden symmetry breaking''
accompanied by finite spin current
at $\q_{\rm sLC}\approx(\pi/2,\pi/2)$.
This sLC order originates from the spin-flipping magnon-exchange process,
called the Aslamazov-Larkin (AL) process
\cite{Onari-SCVC,Onari-FeSe,Yamakawa-FeSe,Chubukov-AL}.
The sLC order is hidden in that neither internal magnetic field 
nor charge density modulation is induced, while
the band-folding by these sLC orders produces the
Fermi arc structure and pseudogap in the DOS
\cite{ARPES-Science2011,Yoshida-arc}.
The derived transition temperature $T_{\rm sLC}$
is higher than that of the stripe-BO, and 
comparable to that of the ferro-BO.
The sLC order will be responsible for 
the pseudogap and electronic nematicity not only in cuprates, 
but also in iridates and $f$-electron systems 
\cite{Sr2IrO4,Sr2IrO4-ARPES,Ikeda,Fujimoto}.

%The sLC is very attractive as the origin of the pseudogap; 
The emergence of the sLC has been discussed 
in various electronic systems 
\cite{Schultz,Ozaki,Nersesyan,Sr2IrO4,Ikeda}.
From the microscopic viewpoint, however,
the mechanism of the sLC is highly nontrivial, 
since the realization condition of the sLC order is
very severe in the extended $U$-$V$-$J$ Hubbard model
within the mean-field theory \cite{Nersesyan}.
In addition, only the case $\q_{\rm sLC}=(\pi,\pi)$ was 
analyzed in previous works.
The present study can explain the 
sLC order based on a simple Hubbard model with on-site $U$,
without assuming the wavevector $\q_{\rm sLC}$.

%%%%%%%%%%%%%%%%%%%%%%%%%%%%%%%%%%%%%%%%%%%%%%%%%%%%%%%%%%%%%%%%%
\section{Spin-fluctuation-driven unconventional orders}
%%%%%%%%%%%%%%%%%%%%%%%%%%%%%%%%%%%%%%%%%%%%%%%%%%%%%%%%%%%%%%%%%

%%%%%%%%%%%%%%%%%%%%%%%%%%%%%%%
\subsection{Model Hamiltonian}
\begin{figure}[htb]
%\vspace{5mm}
\includegraphics[width=.99\linewidth]{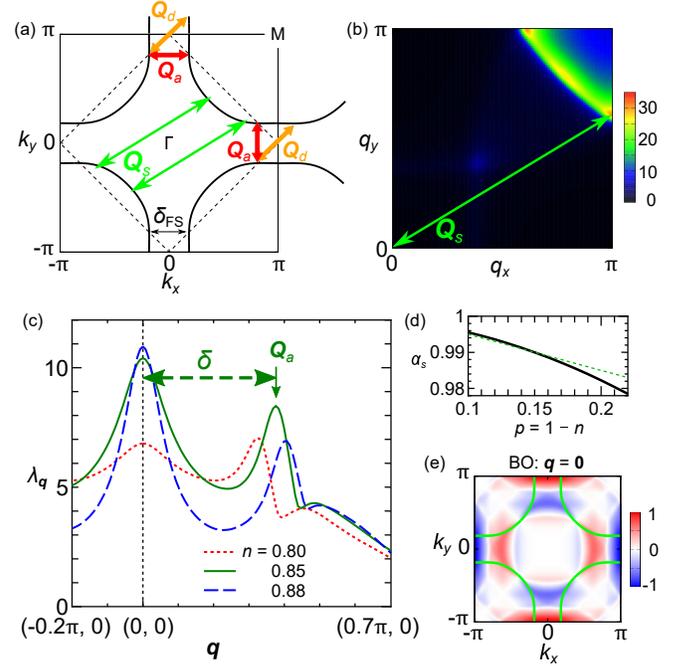}
\caption{
(a) The FS of the present model at $n=0.85$.
$\Q_s$ is the major nesting vectors.
$\Q_{\rm d}=(\delta,\delta)$ and $\Q_{\rm a}=(\delta,0)$
($\delta\approx\delta_{\rm FS}$) are minor nesting vectors.
They correspond to the sLC order wavelength
in the present theory.
(b) $\chi^s(\q)$ given by the RPA.
It shows the incommensurate peak at $\q=\Q_s$.
(c) Obtained eigenvalue $\lambda_\q$ for the BO at $n=0.80\sim0.88$.
They have peaks at $\q={\bm 0}$ and 
$\q=\Q_{\rm a}$.
(d) Relations $\a_S=1-0.444 p^2$ (full line) and
$\a_S=1.01-0.2 p$ (broken line).
(e) Obtained $d$-wave form factor $f_\q(\k)$ for $\q={\bm 0}$
together with the FS.
}
\label{fig:fig2}
\end{figure}

%%%%%%%%%%%%%%%%%%%%%%%%%%%%%%%
Here, we analyze the single-orbital square-lattice Hubbard model 
\begin{eqnarray}
H=\sum_{\k,\s}\e_\k c_{\k\s}^\dagger c_{\k\s}
+U\sum_i n_{i\uparrow}n_{i\downarrow}.
\end{eqnarray}
We denote the hopping integrals $(t_1,t_2,t_3)=(-1,1/6,-1/5)$,
where $t_l$ is the $l$-th nearest hopping integral
\cite{Kontani-ROP,Springer}.
Hereafter, we set the unit of energy as $|t_1|=1$,
which corresponds to $\sim 4000$ [K] in cuprates,
and fix the temperature $T=0.05 \ (\sim 200{\rm K})$.
%$z\ (\sim 1/10)$ is the renormalization factor.
The FS at filling $n=0.85$ is given in Fig. \ref{fig:fig2} (a).
The spin susceptibility
in the random-phase-approximation (RPA) is
$\chi^s(q)= \chi^0(q)/(1-U\chi^0(q))$, where $\chi^0(q)$ 
is the irreducible susceptibility without $U$ and $q\equiv(\q,\w_l)$.
The spin Stoner factor is defined as 
$\a_S\equiv \max_q\{U\chi^0(q)\}= U\chi^0(\Q_s,0)$.
%Here, $\chi^s(q)$ has the highest peak at $\q=\Q_s$.
Figure \ref{fig:fig2} (b) shows the obtained $\chi^s(q)$
at $\a_S=0.99$ ($U=3.27$).
Here, $\chi^s(\Q_s,0)\sim30$ [$1/{t_1}$]
$ \sim80$ [$\mu_{\rm B}^2/{\rm eV}$],
which is still smaller than 
Im$\chi^s(\Q_s,E=31{\rm meV})\sim 200$ [$\mu_{\rm B}^2/{\rm eV}$]
at $T\sim200$K in 60K YBCO
\cite{neutron}.
Thus, $\a_S>0.99$ in real compounds.
Owing to the Mermin-Wagner theorem,
the relation $\a_S\lesssim 1$ is naturally satisfied 
for $U\gg 3.3$ without any fine tuning of $U$
by considering the spin-fluctuation-induced 
self-energy self-consistently
\cite{Kontani-ROP}.

%Various non-Fermi liquid transport phenomena
%are explained by considering the 
%spin-fluctuation-induced VCs
%\cite{Kontani-ROP}.
%Here, we propose the mechanism of sLC order
%based on similar strategy.

%%%%%%%%%%%%%%%%%%%%
%%%%%%%%%%%%%%%%%%%

%%%%%%%%%%%%%%%%%%%%%%%%%%%%%%%%%%%%%%%%%%%%%%%%%%%%%%%%%%%%%%%%%
\subsection{Introduction of singlet and triplet DW equations}
%%%%%%%%%%%%%%%%%%%%%%%%%%%%%%%%%%%%%%%%%%%%%%%%%%%%%%%%%%%%%%%%%

From now on,
we investigate possible exotic density-wave (DW) states
for both charge- and spin-channels with general wavevector ($\q$),
%total spin (singlet or triplet), 
which is generally expressed as
\cite{Nersesyan}
\begin{eqnarray}
D_\q^{\s\rho}(\k)&=&\langle c_{\k_-,\s}^\dagger c_{\k_+,\rho} \rangle
-\langle c_{\k_-,\s}^\dagger c_{\k_+,\rho} \rangle_0 
\nonumber\\
&=& d^c_\q(\k) \delta_{\s,\rho} + \bm{d}^s_{\q}(\k) \cdot \bm{\s}_{\s,\rho}
\end{eqnarray}
where $\k_\pm \equiv \k \pm \q/2$, and 
$d^c_\q(\k)$ $(\bm{d}^s_{\q}(\k))$ 
is the charge (spin) channel order parameter.
It induces the symmetry breaking in the self-energy:
\begin{eqnarray}
\Delta \Sigma_\q^{\s\rho}(\k)
&=& f_\q(\k) \delta_{\s,\rho} + \bm{g}_{\q}(\k) \cdot \bm{\s}_{\s,\rho}
\end{eqnarray}
which we call the form factors in this paper.
Below, we assume $\bm{g}_{\q}(\k)= g_{\q}(\k) {\bm e}_z$ 
without losing generality.
The DW is interpreted as the electron-hole pairing
condensation
\cite{Nersesyan}.

Here, $f_\q(\k)$ is given by the Fourier transformation of 
the spin-independent hopping modulation  
$\sum_{\r_i,\r_j}\delta t_{i,j}e^{i(\r_i-\r_j)\cdot\k} e^{i(\r_i+\r_j)\cdot\q/2}$.
When $\delta t_{i,j}=\pm \delta t_{j,i}$,
the relation $f_\q(\k)= \pm f_\q(-\k)$ holds.
Also, $g_\q(\k)$ is given by the spin-dependent modulation  
$\delta t_{i,j}^\uparrow =-\delta t_{i,j}^\downarrow$.
The even-parity $f_\q(\k)$ and the odd-parity  $g_\q(\k)$
respectively correspond to the BO state and the sLC state.
Both states preserve the time-reversal symmetry.

To find possible DW in an unbiased way, 
we generalize the DW equation \cite{Kawaguchi-CDW}
for both spin/charge channels:
%In order to reveal the possible DW states,
%we solve the following DW equation for $f_\q^\Gamma(\k)$
%in addition to the triplet DW equation  $g_{\q}(\k))$ 
%in later sections:
%
\begin{eqnarray}
& & \!\!\!\! 
\lambda_{\q}f_\q(k)= -\frac{T}{N}\sum_{p}I_\q^c(k,p)G(p_-)G(p_+)f_\q(p) ,
\label{eqn:DWeq3} \\
& & \!\!\!\! 
\eta_{\q}g_\q(k)= -\frac{T}{N}\sum_{p}I_\q^s(k,p)G(p_-)G(p_+)g_\q(p) ,
\label{eqn:DWeq4}
\end{eqnarray}
where 
$\lambda_{\q}$ ($\eta_\q$) is the eigenvalue 
that represents the charge (spin) channel DW instability,
$k\equiv (\k,\e_n)$, $p\equiv (\p,\e_m)$
($\e_n$, $\e_m$ are fermion Matsubara frequencies).
These DW equations are interpreted as the 
``spin/charge channel electron-hole pairing equations''.

The charge (spin) channel kernel function is 
$I_\q^{c(s)} = I_\q^{\uparrow,\uparrow}+(-)I_\q^{\uparrow,\downarrow}$;
$I_\q^{\s,\rho}$ at $\q=0$
is given by the Ward identity $-\delta\Sigma_\s(k)/\delta G_\rho(k')$,
which is composed of 
one single-magnon exchange term and two double-magnon exchange ones:
The former and the latter are called 
the Maki-Thompson (MT) term and the AL terms; 
see Fig. \ref{fig:figS1} in Appendix A.
%Appendix \ref{sec:ApA} for the microscopic derivation.
The lowest order Hartree term $-U\delta_{\s,\rho}$
in $I_\q^{\s,\rho}$ gives the RPA contribution.
while the AL terms are significant for $\a_S\lesssim1$
in various strongy correlated systems 
\cite{Onari-SCVC,Kawaguchi-CDW,Tazai-CeCu2Si2,Tazai-CeB6}.
The significance of the AL processes have been
revealed by the functional-renormalization-group (fRG) study,
in which higher-order vertex corrections are produced in an unbiased way
\cite{Tsuchiizu-CDW,Tsuchiizu-PRL,Tazai-FRG,Tazai-kappa}.
Note that the MT term is important for the 
superconducting gap equation, transport phenomena
\cite{Kontani-ROP},
and cLC order
\cite{Tazai-cLC}.

\begin{figure}[htb]
%\vspace{-10mm}
\includegraphics[width=.99\linewidth]{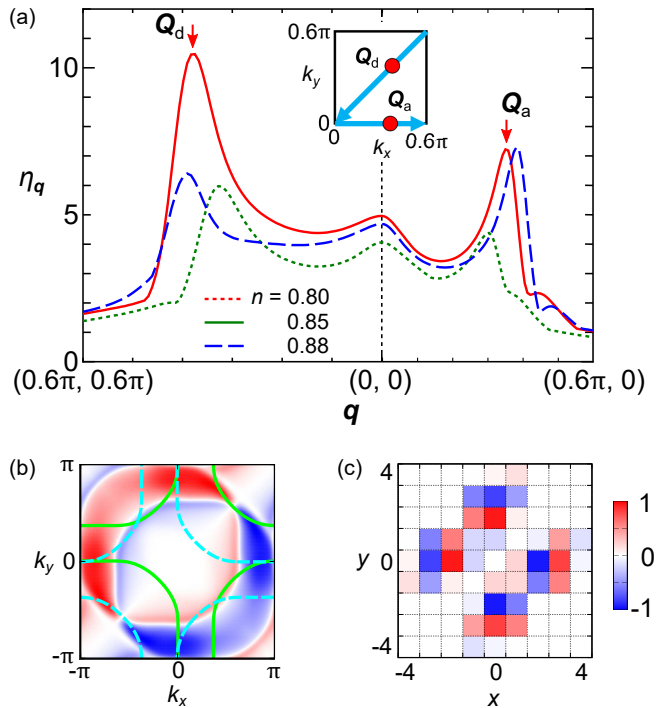}
\caption{
(a) Obtained eigenvalue $\eta_\q$ for spin-channel
DW at $n=0.80 \sim 0.88$.
They have peaks at $\q=\Q_{\rm d}$ and $\Q_{\rm a}$.
(b) Form factor of the diagonal sLC order $g_{\Q_{\rm d}}(\k)$.
We also show the shifted FSs given by $\mu=\e_{\k\pm\Q_{\rm d}/2}$.
(c) ${\rm Im} g_{\Q_{\rm a}}({\bm r})$,
which is even (odd) with respect to $x+y$ ($x-y$).
}
\label{fig:fig3}
\end{figure}

Figure \ref{fig:fig2} (c) shows the charge-channel eigenvalue $\lambda_{\q}$
%for $n=0.80\sim0.88$ 
derived from the DW eq. (\ref{eqn:DWeq3})
\cite{Kawaguchi-CDW,Onari-B2g,Onari-AFB}.
Hereafter, we put $U$ to satisfy the relation
$\a_S=1-0.444 p^2$ with $p\equiv 1-n$,
shown as full line in Fig. \ref{fig:fig2} (d).
The obtained form factor $f_\q(\k)$
at $\q={\bm0},{\Q_{\rm d}}$, shown in Fig. \ref{fig:fig2} (e),
 belongs to $B_{1g}$ symmetry BO, consistently with previous studies
 \cite{Kawaguchi-CDW,Tsuchiizu-CDW}. 
%The present result leads to the successive BO transition scenario,
%that is, ferro-BO at $T^*$ and stripe-BO at $T_{\rm CDW}$
As we discuss in Appendix B,
the large eigenvalue in Fig. \ref{fig:fig2} (c) (and Fig. \ref{fig:fig3} (a))
is strongly suppressed to $O(1)$ by considering
the small quasiparticle weight $z = m/m^* \sim O(10^{-1})$ 
due to the self-energy in cuprates
\cite{Kawaguchi-CDW,Onari-B2g}.

%%%%%%%%%%%%%%%%%%%%
%%%%%%%%%%%%%%%%%%%%%%%%%%%%%%%%%%%%%%%%%%%%

%%%%%%%%%%%%%%%%%%%%%%%%%%%%%%%%%%%%%%%%%%%%%%%%%%%%%
\section{Derivation of sLC order based on triplet DW equation}
%%%%%%%%%%%%%%%%%%%%%%%%%%%%%%%%%%%%%%%%%%%%%%%%%%%%%

%%%%%%%%%%%%%%%%%%%%%%%%%%%%%%%%%%%%%%%%%%%%%%%%%%%%%
\subsection{Origin of sLC order}
%%%%%%%%%%%%%%%%%%%%%%%%%%%%%%%%%%%%%%%%%%%%%%%%%%%%%

Next, we discuss the spin-fluctuation-driven sLC order,
which is the main issue of this manuscript.
%For this purpose, 
%we investigate the spin-triplet DW equation (\ref{eqn:DWeq4}).
Figure \ref{fig:fig3} (a) exhibits the spin-channel eigenvalue $\eta_{\q}$
derived from the DW eq. (\ref{eqn:DWeq4}).
%using $\a_S$ in Fig. \ref{fig:fig2} (b).
Peaks of $\eta_{\q}$ are located at the nesting vectors 
$\q=\Q_{\rm d}$ (diagonal) and $\q=\Q_{\rm a}$ (axial).
The obtained form factor $g_\q(\k)$ at
$\q=\Q_{\rm d}$ (diagonal sLC) is shown in Fig. \ref{fig:fig3} (b).
The odd-parity solution $g_\q(\k)=-g_\q(-\k)$
means the emergence of the sLC order.
The reason for large $\eta_{\Q_{\bm d}}$ is that
all hot spots contribute to the diagonal sLC
as shown in Fig. \ref{fig:fig3} (b).
%and accidental nodes are almost absent.
%These situations make the sLC order eigenvalue larger.
Figure \ref{fig:fig3} (c) shows the form factor in real space
${\rm Im} g_{\Q_{\rm a}}({\bm r})$ with ${\bm r}=(x,y)$.
Here, $\delta t_{i,j}^\s = \s g_{\Q_{\rm a}}({\bm r}_{i-j})
\cos({\bm r}_{i+j}\cdot\Q_{\rm a}/2)$.

To understand why sLC state is obtained,
we simplify Eq. (\ref{eqn:DWeq4}) by taking 
the Matsubara summation analytically 
by approximating that $I_\q^{s}$ and $g_\q(k)$ are static:
\begin{eqnarray}
\eta_{\q}g_\q(\k)= \frac{1}{N}\sum_{\p}I_\q^s(\k,\p)F_\q(\p)g_\q(\p) ,
\label{eqn:DWeq2}
\end{eqnarray}
where 
$\displaystyle 
F_\q(\p)\equiv -T\sum_m G(p_+)G(p_-)
= \frac{n(\e_{\p_+})-n(\e_{\p_-})}{\e_{\p_-}-\e_{\p_+}}$
is a positive function,
and $n(\e)$ is Fermi distribution function;
see Appendix A.
%(Sect. \ref{sec:ApA}).
%When $\q$ is a nesting vector.
%$F_\q(\p)$ is large for wide area of $\p$.
%Therefore, $\eta_{\q}$ show peaks at the nesting vectors.
In general, the peak positions of $\eta_{\q}$
in Eq. (\ref{eqn:DWeq2})
are located at $\q={\bm 0}$ and/or nesting vectors with small wavelength
($\q=\Q_{\rm a}, \Q_{\rm d}$ in the present model).
The reason is that $I_\q \sim T\sum_{p}\chi^s(\p_+)\chi^s(\p_-)$ by AL terms
is large for small $|\q|$,
and $F_\q(\p)$ is large for wide area of $\p$
when $\q$ is a nesting vector.

\begin{figure}[htb]
%\vspace{-10mm}
\includegraphics[width=.99\linewidth]{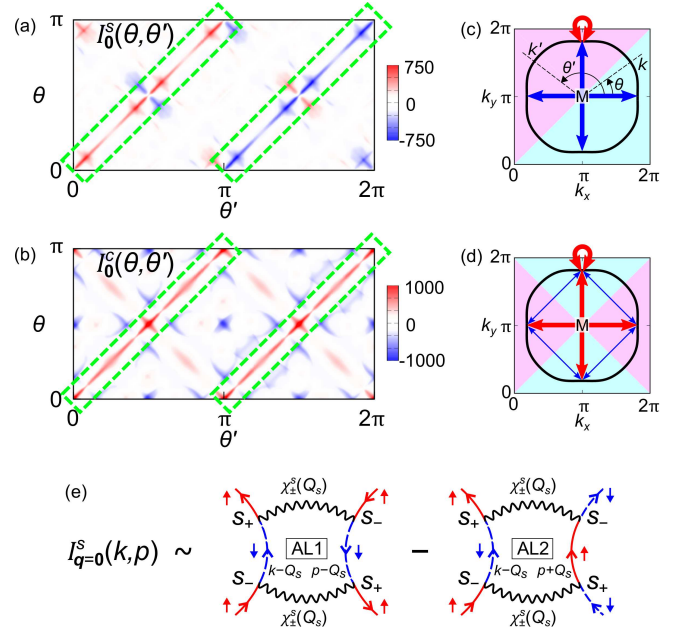}
\caption{
(a) Spin-channel and (b) charge-channel
kernel functions on the FS, $I_{\q={\bm 0}}^{s,c}(\theta,\theta')$,
where $\theta$ represents the position of $\k$.
%; see the next panel.
%$\theta=\{\rm tan}^{-1}(k_y/k_x)$ and
%$\theta'=\{\rm tan}^{-1}(k_y'/k_x')$.
We see that $I_{{\bm 0}}^{s}$ and $I_{{\bm 0}}^{c}$ 
has large negative and positive values
for $\theta'\approx \theta+\pi$, respectively,
due to the p-p channel (= Cooper channel) in AL2.
(c) Origin of $p$-wave sLC order and 
(d) that of $d$-wave BO.
Red (blue) color arrows represent the attractive (repulsive) interaction.
(e) Spin-flipping AL-processes in $I_\q^{s}(\k,\p)$
that give the sLC order.
The wavy lines are transverse spin susceptibilities.
(Spin non-flipping AL processes caused by longitudinal susceptibility 
are cancelled out in $I^{s}=I^{\uparrow,\uparrow}-I^{\uparrow,\downarrow}$.)
The AL1 term with a anti-parallel (p-h) pair 
gives red line on $\theta\approx \theta'$ in (a).
Also, the AL2 term with a parallel (p-p) pair
gives blue line on $\theta\approx \theta'+\pi$ in (a).
}
\label{fig:fig3-2}
\end{figure}

%aaaaaaaaaaaaaaaaaaa
To understand why odd-parity form factor is obtained,
we show the spin-channel ``electron-hole pairing interaction''
$I^s_{\q={\bm 0}}(\k,\k')$ on the FS
in Fig. \ref{fig:fig3-2} (a).
The charge-channel one $I^c_{\q={\bm 0}}(\k,\k')$
is also shown in Fig.\ref{fig:fig3-2} (b).
Here, $\theta$ represents the position of $\k$ 
shown in Figs. \ref{fig:fig3-2} (c) and (d).
$I_\q^s(\k,\k')$ in Fig. \ref{fig:fig3-2} (a)
gives large attractive interaction 
for $\k\approx\k'$ and large repulsive one for $\k\approx-\k'$.
In this case, we naturally obtain $p$-wave form factor $g_\q(\k)$ 
shown in Fig. \ref{fig:fig3} (b),
as we explain in Fig. \ref{fig:fig3-2} (c).
Here, red (blue) arrows represent the attractive (repulsive) interaction.

The strong $\k,\k'$-dependence of $I^s_{\q={\bm 0}}(\k,\k')$
originates from the AL1 and AL2 terms in Fig. \ref{fig:fig3-2} (e), 
or Fig. \ref{fig:figS1} (a) in Appendix A.
Owing to the spin-conservation law, AL terms in
$I^s=I^{\uparrow,\uparrow}-I^{\uparrow,\downarrow}$
originates from the spin-flipping processes due to
transverse spin fluctuations in Fig. \ref{fig:fig3-2} (e),
in proportion to $\chi^{s}_{\pm}(\Q_s)\chi^{s}_{\pm}(\Q_s)$.
(In $I^s$, the spin non-flipping AL processes in proportion to 
$\chi^{s}_{z}(\Q_s)\chi^{s}_{z}(\Q_s)$ are exactly canceled out.)
Therefore, $I^s = \mbox{[AL1]}-\mbox{[AL2]}$.
The AL1 term with the p-h (anti-parallel) pair Green functions
causes large attractive interaction for $\k\approx\k'$,
and the AL2 term with the p-p (parallel) ones does for $\k\approx-\k'$,
as explained in Ref. \cite{Onari-B2g} in detail.
Thus, $\theta,\theta'$-dependence in Fig. \ref{fig:fig3-2} (a) 
and resultant odd-parity solution is understood naturally.

In contrast, the charge channel kernel
$I_\q^c(\k,\k')$ gives an attractive interaction for both
$\k\approx\pm \k'$ as shown in Fig. \ref{fig:fig3-2} (b),
beucase $I^c = 3(\mbox{[AL1]}+\mbox{[AL2]})/2$.
Then, we obtain $d$-wave form factor $f_\q(\k)$ 
in Fig. \ref{fig:fig2} (e),
as we explain in Fig. \ref{fig:fig3-2} (d)
\cite{Kawaguchi-CDW,Onari-B2g}.

In the present transverse spin fluctuation mechanism,
the $\bm{g}$-vector will be parallel to $z$-direction
when $\chi^s_{x(y)}(\Q_s)>\chi^s_z(\Q_s)$ (XY-anisotropy)
due to the spin-orbit interaction (SOI). 
When the XY-anisotropy of $\chi^s_{\mu}(\Q_s)$ is very large,
$I^c$ due to AL terms is multiplied by $2/3$
whereas $I^s$ is unchanged,
so it is suitable condition for the sLC order.

%%%%%%%%%%%%%%%%%%%%
%%%%%%%%%%%%%%%%%%%%%%%%%%%%%%%%%%%%%%%%%%%%

%%%%%%%%%%%%%%%%%%%%%%%%%%%%%%%%%%%%%%%%%%%%%%%%%%%%%%%%%%
\subsection{Filling dependences of sLC/BO instabilities}
%%%%%%%%%%%%%%%%%%%%%%%%%%%%%%%%%%%%%%%%%%%%%%%%%%%%%%%%%%

%%%%%%%%%%%%%%%%%%%%
\begin{figure}[htb]
%\vspace{5mm}
\includegraphics[width=.99\linewidth]{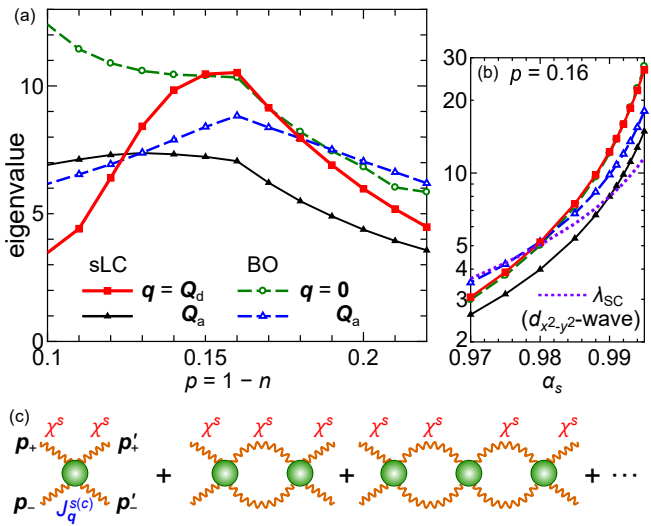}
\caption{
(a) Obtained eigenvalues of sLC and BO as function of $p=1-n$.
(b) $\a_S$-dependences of the eigenvalues at $p=0.16$.
$\lambda_{\rm SC}$ is the eigenvalue of the 
superconducting gap equation.
(c) Diagrammatic expression
for the odd/even parity magnon-pair condensation,
which is the physical origin of the sLC/BO.
% in the present mechanism.
}
\label{fig:fig4}
\end{figure}
%%%%%%%%%%%%%%%%%%%

The sLC and BO eigenvalues are summarized
in Fig. \ref{fig:fig4} (a).
%using $\a_S$ shown in Fig. \ref{fig:fig2} (d).
The relation $\eta_{\Q_{\rm d}}>\lambda_{\Q_{\rm a}}$ 
around the optimal doping ($p\sim0.15$) means that 
the sLC transition temperature $T_{\rm sLC}$ is
higher than $T_{\rm CDW}$, as in Fig. \ref{fig:fig1} (a).
%The obtained result is robust in that
%similar numerical results are obtained based on the $d$-$p$ orbital Hubbard model
%introduced in Ref. \cite{Tsuchiizu-CDW,Kawaguchi-CDW}.
The robustness of Fig. \ref{fig:fig4} (a) is verified 
in Appendix B.
% (Sec. \ref{sec:ApB}).
%based on another condition $\a_S=1.01-0.2p$,
%which is shown as a broken line in Fig. \ref{fig:fig2} (d),  
We also verify in Fig. \ref{fig:fig4} (b) that
both $\eta_{\Q_{\rm d}}$ and $\lambda_{\Q_{\rm a}}$
are larger than the $d_{x^2-y^2}$-wave
superconducting eigenvalue $\lambda_{\rm SC}$ 
for $\a_S\gtrsim0.98$.
% at $p=0.16$.
Here, $\lambda_{\rm SC}$ is derived from the gap equation 
\begin{eqnarray}
\lambda_{\rm SC} \Delta(k)= T\sum_{p}V^{\rm SC}(k,p)|G(p)|^2\Delta(p),
\label{eqn:gap-eq}
\end{eqnarray}
%
%$\lambda_{\rm SC} \Delta(k)= T\sum_{p}V^{\rm SC}(k,p)|G(p)|^2\Delta(p)$,
where $V^{\rm SC}(k,p)=-\frac32 U^2\chi^s(k-p)+\frac12 U^2\chi^c(k-p)-U$
is the MT-type kernel \cite{d-wave}.
Note that $\eta_\q, \lambda_\q < \lambda_{\rm SC}$
if AL terms are dropped \cite{Norman}.
The large eigenvalues in Fig. \ref{fig:fig4}
are suppressed to $O(1)$ by small $z$; see 
Refs. \cite{Kawaguchi-CDW,Onari-B2g,Kontani-ROP}
and Appendix B. 
%As shown in Figs. \ref{fig:fig4} (a) and (b),
%the sLC eigenvalue is as large as the BO eigenvalue for wide doping range.
%This result means that the sLC order 
%and the ferro-BO occur at almost the same temperatures at $T\sim T^*$.
We stress that sLC is not suppressed by the
ferro-BO that induces neither Fermi arc nor pseudogap,
as explained in Appendix B.
% (Sec. \ref{sec:ApB}).
%The Dirac point inside the Fermi arc due to $p$-wave $g_{\q}(\k)$ 
%would disappear when the $d$-wave BO sets in below $T_{\rm CDW}$.

Here, we analyzed
the sLC/BO in terms of the electron-hole pairing.
Another physical interpretation of the sLC/BO is the 
``condensation of odd/even parity magnon-pairs'',
which is the origin of the nematic order in quantum spin systems
\cite{Andreev,Coleman,Shannon}.
In fact, the two-magnon propagator shown in Fig. \ref{fig:fig4} (c) 
diverges when the eigenvalue of DW equation reaches unity, 
as we explain in Appendix C.
% (Sec. \ref{sec:ApC}).
That is, triplet (singlet) magnon-pair condensation occurs 
at $T=T_{\rm sLC}(T_{\rm CDW})$.
Thus, the sLC/BO discussed here and 
the spin nematic order in quantum spin systems
are essentially the same phenomenon.
%Thus, clear physical picture of the sLC/BO
%is achieved by the present weak-coupling theory,
%in which the instability of short-lived quasiparticles is focused on.

%%%%%%%%%%%%%%%%%%%%%%%%%%%%%%%%%%%%%%%%%%%%%%%%%%%%%%
\subsection{Fermi arc and pseudogap under sLC order}
%%%%%%%%%%%%%%%%%%%%%%%%%%%%%%%%%%%%%%%%%%%%%%%%%%%%%%

\begin{figure}[htb]
%\vspace{5mm}
\includegraphics[width=.99\linewidth]{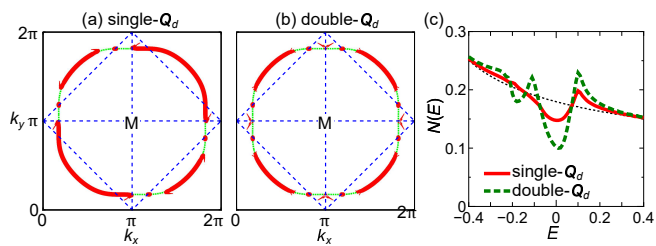}
\caption{
(a) Fermi arc structure due to the single-$\q$ order,
(b) that due to the double-$\q$ order, and
(c) pseudogap in the DOS
due to the diagonal sLC order ($\q={\bm Q}_{\rm d}$).
%and the (d)-(f) axial sLC order ($\q={\bm Q}_{\rm a}$).
%In both cases, we set $g^{\rm max}=0.1$.
%
%(d) Obtained spin current from the center site (A-D) 
%to different sites under the diagonal sLC with period $4 a_{\rm Cu-Cu}$.
%The real space pattern is depicted in 
%Fig. \ref{fig:fig1} (c), with sites A-D in a unit cell.
%(e)-(g) Fermi arc structures and pseudogap 
%due to the axial sLC order with $\q={\bm Q}_{\rm a}$.
In calculating (a)-(c), we introduced BCS-type cutoff energy 
$\w_c=0.5$ for the band-hybridization by $g_\q(\k)$.
}
\label{fig:fig5}
\end{figure}

Now, we discuss the band-folding and hybridization gap
due to the diagonal sLC order.
Figures \ref{fig:fig5} (a) and (b) show the Fermi arc structures
%induced by  sLC ($\q_{\rm sLC}=\Q_{\rm d}$) order
in the cases of (a) single-$\q$ and (b) double-$\q$ orders.
%(The latter state is given by the coexistence of
%$\Q_{\rm a}=(\delta,0)$ and $\Q_{\rm a}'=(0,\delta)$ sLC order parameters.)
We set $g^{\rm max}\equiv \max_\k \{g_{\Q_{\rm d}}(\k)\} =0.1$.
Here, the folded band structure under the sLC order with 
finite $\q_{\rm sLC}$ is ``unfolded''
into the original Brillouin zone by following Ref. \cite{Ku}
to make a comparison with ARPES results.
The Fermi arc due to the single-$\q$ order
in Fig. \ref{fig:fig5} (a) belongs to $B_{2g}$ symmetry.
%reflection symmetry about $k_x$- and $k_y$-axes.
In contrast, the Fermi arc due to the double-$\q$ order
in Fig. \ref{fig:fig5} (b) preserves the $C_4$ symmetry.
The resultant pseudogap in the DOS is shown in Fig. \ref{fig:fig5} (c).
The unfolded band structure in the single-$\Q_{\rm d}$ sLC order
is displayed in Fig. \ref{fig:figS2} in Appendix B.
%ix \ref{sec:ApB}.
% \cite{SM}.
%The unfolded band structure is shown in Fig. S2 in the SM: B.

%%%%%%%%%%%%%%%%%%%%
%%%%%%%%%%%%%%%%%%%

Recent magnetic torque measurements revealed the 
$B_{1g}$ symmetry breaking at $T=T^*$ occurs in YBCO
\cite{Y-Sato},
while $B_{2g}$ one appears in Hg-based cuprates
\cite{Hg-Murayama}.
To understand different symmetry breakings,
we examine the $t_3$ dependence of DW equation solution in Appendix B 2:
As shown in Figs. \ref{fig:phase-t3} (b)-(e),
the sLC wavevector $\q_{\rm sLC}$ changes from $\Q_{\rm d}$
to $\Q_{\rm a}$ for larger $t_3/t_1$.
When $\q_{\rm sLC}=\Q_{\rm d}$,
the symmetry of the Fermi arc is $B_{2g}$ in Fig. \ref{fig:fig5} (a).
On the other hand, 
the Fermi arc has $B_{2g}$ symmetry 
in the axial sLC order in Fig. \ref{fig:fig1} (d),
as we show in Fig. \ref{fig:axial-sLC} (b).

Since the van Vleck susceptibility becomes anisotropic
when $C_4$ symmetry of the FS is broken
\cite{Kawaguchi-B2g},
the reported compound-dependent symmetry breaking 
\cite{Y-Sato,Hg-Murayama}
would be explained by the sLC order scenario.
This is an important future issue.

%%%%%%%%%%%%%%%%%%%%%%%%%%%%%%%%%%%%%%%%%%%%%%%%%%%%%%
\subsection{Spin-current pattern under sLC order}
%%%%%%%%%%%%%%%%%%%%%%%%%%%%%%%%%%%%%%%%%%%%%%%%%%%%%%

Next, we investigate the spin current in real space,
which is driven by a fictitious Peierls phase 
due to the ``spin-dependent self-energy'' $\delta t_{i,j}^\s = \s g_{i,j}$.
As shown in Fig. \ref{fig:fig3} (c), $\delta t_{i,j}$ is purely imaginary
and odd with respect to $i \leftrightarrow j$.
%Here, $g_{i,j}$ is the Fourier transformation of the form factor $g_{\q}(\k)$.
%Since $g_{\q}(\k)$ is odd-function, $\delta t_{i,j}^\s$ is purely imaginary.
%and the relation $\delta t_{i,j}^\uparrow=-\delta t_{j,i}^\downarrow$ holds.
The conservation law $\dot{n}_i^\s = \sum_j j_{i,j}^\s$
directly leads to the definition the spin current operator 
from site $j$ to site $i$ as
$j_{i,j}^\s =-i\sum_{\s}\s (h_{i,j}^\s c_{i\s}^\dagger c_{j\s}-(i \leftrightarrow j))$,
where $h_{i,j}^\s = t_{i,j}+\delta t_{i,j}^\s$.
Then, the spontaneous spin current from $j$ to $i$ is
$J_{i,j}^s=\langle j_{i,j}^s \rangle_{\hat{h}^\s}$.

Here, we calculate the spin current for the
commensurate sLC order at $\q_{\rm sLC}=(\pi/2,\pi/2)$,
which is achieved by putting $n=1.0$.
% (The obtained form factor is very similar to Fig. \ref{fig:fig3} (b).)
Then, the unit cell are composed four sites A-D.
In Fig. \ref{fig:S-current} in Appendix D,
we show the obtained spin current $J_{i,j}^s$ 
from the center site ($j={\rm A}$-${\rm B}$) to different site 
in Fig. \ref{fig:fig1} (c).
%, by setting $g^{\rm max}=0.1$.
%The obtained current is $|J_{i,j}^s|\sim10^{-2}$ in magnitude in unit $|t_1|/\hbar$.
The derived spin current pattern 
between the nearest and second-nearest sites
is depicted in Fig. \ref{fig:fig1} (c).
%The spin current is exactly conserved at each site.

The Fermi arc structure and the DOS 
in Fig. \ref{fig:fig5} are independent of
the phase shift $g_\q \rightarrow e^{i\psi}g_\q$.
In contrast, the real space current pattern 
depends on the phase shift.
We discuss other possible sLC patterns 
in Appendix D.
% (Sect. \ref{sec:ApD}).
The charge modulation due to the sLC is 
%proportional to $(g^{\rm max})^2$,
just $|\Delta n_i|\sim 5\times10^{-4}$ for $g^{\rm max}=0.1$
since $|\Delta n_i| \propto (g^{\rm max})^2$.
Thus, experimental detection of translational symmetry breaking 
by sLC order may be difficult.
However, the cLC is induced by applying uniform magnetic field 
parallel to ${\bm g}_{i,j}$.
In the present sLC state, under 10 T magnetic field,
the induced cLC gives $\Delta H\sim \pm 0.1$ Oe when $m^*/m\sim10$, 
which may be measurable by NMR or $\mu$SR study.

%Thus, the sLC state can be confirmed experimentally 
%by measuring the field-induced staggered magnetic field

%%%%%%%%%%%%%%%%%%%%%%%%%%%%%%%%%%%%%%%%%%%%%%%%%%%%%%
\subsection{sLC fluctuation mediated superconductivity}
%%%%%%%%%%%%%%%%%%%%%%%%%%%%%%%%%%%%%%%%%%%%%%%%%%%%%%
%SC
Finally, we discuss that the sLC fluctuations 
can contribute to the $d_{x^2-y^2}$-wave pairing mechanism.
The sLC fluctuations 
connects the close hot spots at $\bm{P}$ and $\bm{P'}=\bm{P}+\q$
in Fig. \ref{fig:fig2} (a).
At both points, the $d_{x^2-y^2}$-wave gap function $\Delta_\k$ has the same sign. 
The pairing interaction mediated by the spin-channel sLC fluctuations 
between singlet pairs $(\bm{P},-\bm{P})$ and $(\bm{P'},-\bm{P'})$ is
\begin{eqnarray}
V^{\rm sLC}(\bm{P},\bm{P'}) \propto g_{\q}(\k)g_{\q}(-\k)\cdot(-\chi_{\rm sLC}(\q)) ,
\label{eqn:VSC}
\end{eqnarray}
where $\k=(\bm{P}+\bm{P'})/2$ and $\q=\bm{P}-\bm{P'}\sim \Q_{\rm a}$.
$\chi_{\rm sLC}(\q) \ (>0)$ is the sLC susceptibility and 
$g_{\q}(\k)=-g_{\q}(-\k)$ is its form factor.
%since $g_{\q}(\k)$ is real and odd-function.
Thus, Eq. (\ref{eqn:VSC}) give positive (=attractive) pairing 
interaction between close hot spots.
%, and therefore the $d_{x^2-y^2}$-wave 
%Note that $g_{\q}(\k)$ is real, and the minus sign in front of $\chi_{\rm sLC}$ 
%in Eq. (\ref{eqn:VSC}) originates from the fact that the sLC fluctuations 
%are spin-channel.
%In the case of conventional spin fluctuations with $g={\rm const}$,
%the pairing interaction in Eq. (\ref{eqn:VSC}) is negative (=repulsive).
%In contrast, in the case of the sLC fluctuations, 
%Eq. (\ref{eqn:VSC}) is positive because of the add-parity relation
%$g_{\q}(\k)=-g_{\q}(-\k)$.
%Thus, the sLC fluctuations give positive (=attractive) pairing 
%interaction between close hot spots, and therefore the $d_{x^2-y^2}$-wave 
%$T_{\rm c}$ can be enlarged by the odd-parity sLC fluctuations.
The derivation of Eq. (\ref{eqn:VSC}) is given in Appendix A
and in SM F of Ref. \cite{Onari-AFB}.
This mechanism may be important for slightly over-doped cuprates
with $T_{\rm sLC}\lesssim T_{\rm c}$.

%%%%%%%%%%%%%%%%%%%%%%%%%%%%%%%%
\section{Summary}
%%%%%%%%%%%%%%%%%%%%%%%%%%%%%%%%

In summary
we proposed a novel and long-sought formation mechanism for 
the nanoscale spin-current order, which 
violates the parity and translational symmetry
while time-reversal symmetry is preserved.
It was revealed that 
the formation of triplet odd-parity electron-hole pairs
that is mediated by spin fluctuations, 
and therefore the spontaneous sLC is established at $T=T_{\rm sLC}$.
In the present spin-fluctuation mechanism,
the condensation of odd-parity magnon-pairs occurs simultaneously.
%The real-space sLC orders
%are shown in Fig. \ref{fig:fig1} (c).
The band-folding by the sLC orders results in the formations of 
the Fermi arc structure and pseudogap at $T \sim T^*$.
In the sLC state, a staggered moment is expected to appear
under the uniform magnetic field.
The sLC order will be a key ingredient in understanding pseudogap phase 
and electronic nematicity not only in cuprates,
but also in iridates \cite{Sr2IrO4,Sr2IrO4-ARPES} and
heavy-fermion compound.
It is an important future issue to 
incorporate the self-energy effect into the present theory.

\acknowledgements
The authors are grateful to Y. Matsuda, 
K. Yamada, T. Moriya, and A. Kobayashi
for the fruitful comments and discussions.
This work was supported by the ``Quantum Liquid Crystals''
No. JP19H05825 KAKENHI on Innovative Areas from JSPS of Japan, 
and JSPS KAKENHI (JP18H01175, JP20K22328, JP20K03858, JP17K05543).

\appendix

%%%%%%%%%%%%%%%%%%%%%%%%%%%%%%%%%%%%%%%%%%
\section{Derivation of singlet and triplet DW equations}
\label{sec:ApA}
% bound state of magnons: BO, sLC, 

\begin{figure}[htb]
%\vspace{5mm}
\includegraphics[width=.99\linewidth]{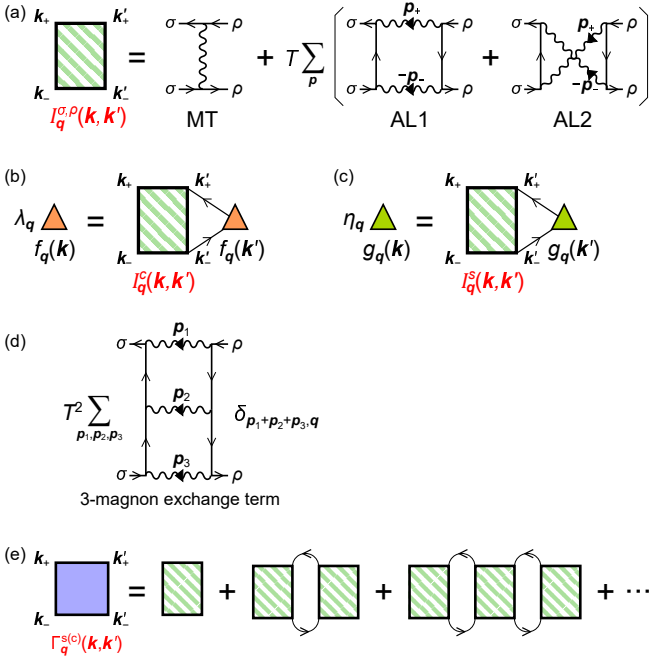}
\caption{
(a) Irreducible four-point vertex $I_\q^{\s,\rho}(k,k')$
composed of one MT term and two AL terms.
(b) Linearized singlet DW equation
with the kernel $I^{c}\equiv I^{\s,\s}+I^{\s,-\s}$.
(c) Linearized triplet DW equation
with the kernel $I^{s}\equiv I^{\s,\s}-I^{\s,-\s}$.
(d) A three-magnon exchange term, which is less important.
(e) Full four-point vertex function $\Gamma^{s(c)}_\q(k,k')$
given by solving the DW equation.
The sLC order (BO) emerges when $\Gamma^{s(c)}_\q(k,k')$ diverges.
}
\label{fig:figS1}
\end{figure}

Here, we discuss the linearized density-wave (DW) equation
driven by spin fluctuations.
For this purpose, 
we introduce the irreducible four-point vertex function 
$I_\q^{\s,\rho}(k,k')$.
It is given by the Ward identity at $\q=0$, that is,
$I_\q^{\s,\rho}(k,k') \equiv -\delta\Sigma_\s(k)/\delta G_{\rho}(k')$.
Here, we use the one-loop self-energy given as
\begin{eqnarray}
&&\!\!\!\!\!\!\!\!
\Sigma_\s(k)= T\sum_q U^2 \chi_{\rm L}^\s(q)G_\s(k+q)
\nonumber \\
&&
+ T\sum_p U^2 (\chi_{\rm T}^\s(q)-\chi_{\rm T}^{(0)\s}(q))G_{-\s}(k+q) ,
\end{eqnarray}
where $\chi_{\rm L}^\s(q)$ and $\chi_{\rm T}^\s(q)$
are longitudinal and transverse susceptibilities.
They are given as
\begin{eqnarray}
\chi_{\rm L}^{\s}(q)&=& 
\chi_{\rm L}^{(0)\s}(q)
(1-U^2\chi_{\rm L}^{(0)\s}(q)\chi_{\rm L}^{(0)-\s}(q))^{-1},
\\
\chi_{\rm T}^\s(q)&=& \chi_{\rm T}^{(0)\s}(q)(1-U\chi_{\rm T}^{(0)\s}(q))^{-1},
\end{eqnarray}
where
$\chi_{\rm L}^{(0)\s}(q)= -T\sum_p G_{\s}(p)G_{\s}(p+q)$
and 
$\chi_{\rm T}^{(0)\s}(q)= -T\sum_p G_{\s}(p)G_{-\s}(p+q)$
are longitudinal and transverse irreducible susceptibilities.
Then, the irreducible vertex function 
$I_\q^{\s,\rho}(k,k')$ given by the Ward identity
is composed of one MT term and two AL terms in Fig. \ref{fig:figS1} (a).
Note that $I_\q^{\s,\rho}$ contains  
the lowest order Hartree term $-U\delta_{\s,\rho}$.

First, we derive the charge-channel (singlet) DW equation
in the absence of the magnetic field,
where the form factor is independent of spin:
$f_\q^\uparrow(k) =f_\q^\downarrow(k) = f_\q(k)$.
The singlet DW equation was
introduced in the study of Fe-based superconductors
\cite{Onari-B2g}
and cuprate superconductors
\cite{Kawaguchi-CDW}.
It is given as
\begin{eqnarray}
\lambda_{\q}f_\q(k)&=& -\frac{T}{N}\sum_{k'}I_\q^c(k,k')G(k_-')G(k_+')f_\q(k') ,
% \nonumber  \\
%&=& \frac{T}{N}\sum_{k'}K_\p(k,k')f_\p(k')
\label{eqn:DWeq1S}
\end{eqnarray}
which is shown in Fig. \ref{fig:figS1} (b), and 
$\k_\pm \equiv \k\pm \q/2$.
Here,
$I_\q^c(k,k')=I_\q^{\s,\s}(k,k')+I_\q^{\s,-\s}(k,k')$.
It is given as
\begin{eqnarray}
& &\!\!\!\!\!\!\!\!\!\!
I_{\bm{q}}^c(k,k')=
-\frac{3}{2} V^{s}(k-k')-\frac{1}{2} V^{c}(k-k') 
\nonumber \\
& & 
+\frac{T}{N}\sum_{p}
 [\frac{3}{2} V^{s}(p_+)V^{s}(p_-)+\frac{1}{2} V^{c}(p_+)V^{c}(p_-)]
\nonumber \\
& &\ \ \ \ \ \times G(k-p)G(k'-p)
\nonumber \\
& & 
+\frac{T}{N}\sum_{p}
 [\frac{3}{2} V^{s}(p_+)V^{s}(p_-)+\frac{1}{2} V^{c}(p_+)V^{c}(p_-)]
\nonumber \\
& &\ \ \ \ \ \times G(k-p)G(k'+p),
%-({\rm Double\;counting\;} [\hat{\Gamma}^{s(c)}]^2 \;{\rm terms})
\label{eqn:IcS} 
\end{eqnarray}
where $p=(\p,\w_l)$, 
$\hat{V}^{s}(q)=U+U^2\hat{\chi}^{s}(q)$,
and 
$\hat{V}^{c}(q)=-U+U^2\hat{\chi}^{c}(q)$.
The first, the second, the third terms in Eq. (\ref{eqn:IcS})
corresponds to the MT, AL1 and AL2 terms in Fig. \ref{fig:figS1} (a).

In cuprates, Eq. (\ref{eqn:DWeq1S}) gives even-parity solution 
with wavevector $\q={\bm 0}$ and $\q\approx (\pi/2,0)$.
This singlet and even-parity electron-hole condensation
is interpreted as the BO.

%%%%%%%%%%%%%%%%%%%%
%%%%%%%%%%%%%%%%%%%

Next, we derive the spin-channel (triplet) DW equation
in the absence of the magnetic field,
the spin-dependent form factor is 
$g_\q(k)\equiv g_\q^\uparrow(k) =-g_\q^\downarrow(k)$.
It is given as
\begin{eqnarray}
\eta_{\q}g_\q(k)&=& -\frac{T}{N}\sum_{k'}I_\q^s(k,k')G(k_-')G(k_+')g_\q(k') ,
% \nonumber  \\
%&=& \frac{T}{N}\sum_{k'}K_\p(k,k')f_\p(k')
\label{eqn:DWeq2S}
\end{eqnarray}
which is shown in Fig. \ref{fig:figS1} (c).
Here,
$I_\q^s(k,k')=I_\q^{\s,\s}(k,k')-I_\q^{\s,-\s}(k,k')$.
It is given as
\begin{eqnarray}
& &\!\!\!\!\!\!\!\!\!\!
I_{\bm{q}}^s(k,k')=
\frac{1}{2} V^{s}(k-k')-\frac{1}{2} V^{c}(k-k')
\nonumber \\
& & 
+\frac{T}{N}\sum_{p}
 [V^{s}(p_+)V^{s}(p_-)+\frac{1}{2} V^{s}(p_+)V^{c}(p_-)
\nonumber \\
& &\ \ \ \ \ +\frac{1}{2} V^{c}(p_+)V^{s}(p_-)] G(k-p)G(k'-p)
\nonumber \\
& & 
+\frac{T}{N}\sum_{p}
 [-V^{s}(p_+)V^{s}(p_-)+\frac{1}{2} V^{s}(p_+)V^{c}(p_-)
\nonumber \\
& &\ \ \ \ \ +\frac{1}{2} V^{c}(p_+)V^{s}(p_-)] G(k-p)G(k'+p) ,
%-({\rm Double\;counting\;} [\hat{\Gamma}^{s(c)}]^2 \;{\rm terms})
\label{eqn:IsS} 
\end{eqnarray}
where the first, the second, the third terms in Eq. (\ref{eqn:IcS})
corresponds to the MT, AL1 and AL2 terms in Fig. \ref{fig:figS1} (a).
The AL terms with $V^{s}(p_+)V^{s}(p_-)$ 
are shown in Fig. \ref{fig:fig3-2} (e).
In cuprates, Eq. (\ref{eqn:DWeq2S}) gives the 
odd-parity solution at wavevector $\q=(\pi/2,\pi/2)$ and $(\pi/2,0)$.
This triplet and odd-parity electron-hole pairing
is interpreted as the spin-loop-current (sLC).

In both Eqs. (\ref{eqn:IcS}) and (\ref{eqn:IsS}),
the AL terms are proportional to 
$\phi^{(2)}_\q\equiv T\sum_{\p_1,\p_2} V^{s}(\p_1)V^{s}(\p_2)\cdot
\delta_{\p_1+\p_2,\q}$.
The AL terms are significant
when the spin fluctuations are large,
since both $V^{s}(\p_1)$ and $V^{s}(\q-\p_1)$ take large value 
simultaneously when $\p_1\approx \Q_s$ in the case of $\q\approx{\bm0}$.
If we put $V^{s}(\p)\propto \xi^2/(1+\xi^2(\p-\Q_s)^2)$
at zero Matsubara frequency,
where $\xi\ (\gg1)$ is the magnetic correlation length,
$\phi^{(2)}_{\q={\bm 0}} \propto T \xi^2$ in two-dimensional systems.
Therefore, double-magnon exchange (AL) terms induce not only BO,
but also the sLC order when $\xi\gg1$.
A three-magnon exchange term shown
in Fig. \ref{fig:figS1} (d) 
is proportional to $\phi^{(3)}_\q \equiv T^2\sum_{\p_1,\p_2,\p_3} 
V^{s}(\p_1)V^{s}(\p_2)V^{s}(\p_3)\cdot\delta_{\p_1+\p_2+\p_3,\q}$.
Then, $\phi^{(3)}_{\q={\bm0}} \propto T^2 \xi^2$ in two-dimensional 
systems for $\q\sim \Q_s$,
which is smaller than $\phi^{(2)}_{\q={\bm0}}$ 
at low temperatures $T\ll E_{\rm F}$.
Thus, the AL process would be the most significant,
which is also indicated by functional-renormalization-group studies
\cite{Tsuchiizu-CDW}.

The electron-hole pairing order
% with time-reversal symmetry are 
is generally expressed in real space as follows
\cite{Nersesyan}:
\begin{eqnarray}
D_{i,j}^{\s,\rho} &\equiv& \langle c_{i\s}^\dagger c_{j\rho} \rangle
-\langle c_{i\s}^\dagger c_{j\rho} \rangle_0
\nonumber \\
&=& d^c_{i,j} \delta_{\s,\rho} + \bm{d}^s_{i,j} \cdot \bm{\s}_{\s,\rho},
\label{eqn:D1S}
\end{eqnarray}
where $D_{i,j}^{\s,\rho}=\{D_{j,i}^{\rho,\s}\}^*$,
and $d^c_{i,j}$ (${\bm d}^s_{i,j}$) is spin singlet (triplet) pairing.
It induces the symmetry breaking in the self-energy:
\begin{eqnarray}
\Delta \Sigma_{i,j}^{\s\rho}
&=& f_{i,j} \delta_{\s,\rho} + \bm{g}_{i,j} \cdot \bm{\s}_{\s,\rho}
\end{eqnarray}
which we call the form factors in this paper.
The BO is given by real even-parity function $f_{i,j}=f_{j,i}$, and
the sLC is given by pure imaginary odd-parity vector 
${\bm g}_{i,j}=-{\bm g}_{j,i}$.
Both orders preserve the time-reversal symmetry.
%The forlatter is a spin-triplet electron-hole pairing state
%\cite{Nersesyan}.
Note that $f_\q(k)$ and $g_\q(k)$
in Eqs. (\ref{eqn:DWeq1S}) and (\ref{eqn:DWeq2S})
correspond to $f_{i,j}$ and $g_{i,j}^z$,
respectively.
%In the main text, we choose $\bm{g}\parallel \bm{e}_z$.

Finally, we discuss the effective interaction driven by 
the BO/sLC fluctuations.
By solving the DW equation (\ref{eqn:DWeq1S}), we obtain the 
full four-point vertex function $\Gamma^{c}_\q(k,k')$
that is composed of $I_\q^{c}$ and $G(k_+)G(k_-)$
shown in Fig. \ref{fig:figS1} (e),
which increases in proportion to $(1-\eta_\q)^{-1}$.
%If we solve Eq. (\ref{eqn:DWeq2S}),
%we obtain $\Gamma^{s}_\q(k,k')$ composed of $I_\q^{s}$ and $G(k_+)G(k_-)$,
%which increases in proportion to $(1-\lambda_\q)^{-1}$.
Thus, we obtain the relation
$\Gamma^{c}_\q(k,k')\approx f_\q(k)\{f_\q(k')\}^* {\bar I}_\q^c(1-\lambda_\q)^{-1}$,
which is well satisfied when $\lambda_\q$ is close to unity.
Here, 
${\bar I}_\q^{c(s)}\equiv T^2\sum_{k,k'} \{f_\q(k)\}^* I_\q^{c(s)}(k,k')f_\q(k')/T\sum_{k} |f_\q(k)|^2$.
In the same way, we obtain the relation
$\Gamma^{s}_\q(k,k')\approx g_\q(k)\{g_\q(k')\}^* {\bar I}_\q^s(1-\eta_\q)^{-1}$.
Thus, it is apparent that the sLC order $g$ (BO $f$) 
emerges when $\Gamma^{s(c)}_\q(k,k')$ diverges.

The pairing interaction due to the sLC fluctuations is
given by the full four-point vertex.
It is approximately expressed as
$V^{\rm SC}(\k_+,\k_-)= -\Gamma_\q^s(\k,-\k) 
\propto -g_{\q}(\k)\{g_{\q}(-\k)\}^*(1-\eta_\q)^{-1}$. 
Since $g$ is odd-function, the sLC fluctuations
cause attractive interaction:
$V^{\rm SC}(\k_+,\k_-) \propto -|g_{\q}(\k)|^2(1-\eta_\q)^{-1}$.

%%%%%%%%%%%%%%%%%%%%%%%%%%%%%%%%%%%%%%%%%%
\section{Additional numerical results of DW equations}
\label{sec:ApB}

\subsection{Enhancement of the sLC instability
under the finite ferro BO}

In Figs. \ref{fig:fig4} (a) and (b) in the main text,
the sLC eigenvalue $\eta_{\q=\Q_{\rm d}}$ is comparable to 
the BO eigenvalue $\lambda_{\q=\bm{0}}$ for a wide doping range.
This result means that the sLC order at $\q_{\rm sLC}=\Q_{\rm d}$
and the ferro-BO occur at almost the same temperature $\sim T^*$.
Here, we discuss the possibility of coexistence 
of sLC order and ferro-BO.

%%%%%%%%%%%%%%%%%%%%
\begin{figure}[htb]
%\vspace{5mm}
\includegraphics[width=.8\linewidth]{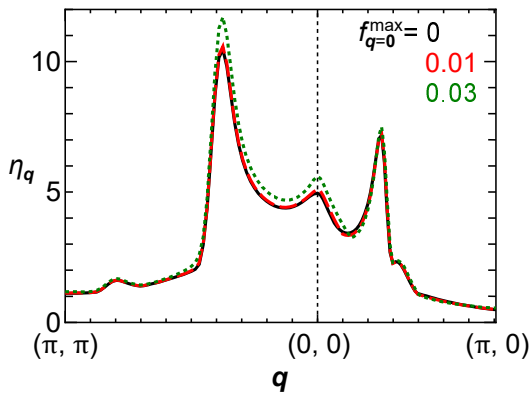}
\caption{
Obtained $\eta_\q$ for $n=0.85$ ($\a_S=0.99$) under the ferro-BO with 
$f_{\q=\bm{0}}^{\rm max}=0,\ 0.01, \ 0.03$.
Thus, the ferro-BO does not prohibit the emergence of the sLC order.
}
\label{fig:figS7}
\end{figure}
%%%%%%%%%%%%%%%%%%%

Since the ferro-BO does not induce the band-folding and pseudogap,
the sLC order will emerge even if
the ferro-BO transiting temperature is higher.
To verify this expectation,
we calculated the triplet DW equation (\ref{eqn:DWeq2S})
under the ferro-BO with $f_{\q=\bm{0}}^{\rm max}=0,\ 0.01, \ 0.03$.
Figure \ref{fig:figS7} shows the
eigenvalue of sLC as function of $\q$
for $n=0.85$ and $U=3.27$ ($\a_S=0.99$)
under the ferro-BO form factor obtained by the 
spin-singlet DW equation (\ref{eqn:DWeq1S}).
It is verified that
the ferro-BO does not prohibit the emergence of the sLC order.
The eigenvalue $\eta_\q$ slightly increases with $f_{\q=\bm{0}}^{\rm max}$,
since the spin Stoner factor $\a_S$ is enlarged by the ferro-BO
\cite{Kawaguchi-CDW,Tsuchiizu-CDW}.

\subsection{Change in the phase diagram by $t_3$}

In the main text, we show that the sLC eigenvalue 
at $\q=\Q_{\rm d}$ develops as large as 
the ferro-BO eigenvalue near the optimally-doping case ($p\sim0.15$),
based on the Hubbard model with the hopping integrals 
$(t_1,t_2,t_3)=(-1.0,1/6,-1/5)$.
The obtained Fermi surface (FS) has the flat part near the 
Brillouin zone boundary, which captures the characteristic of YBCO compounds.

\begin{figure}[htb]
%\vspace{5mm}
\includegraphics[width=.99\linewidth]{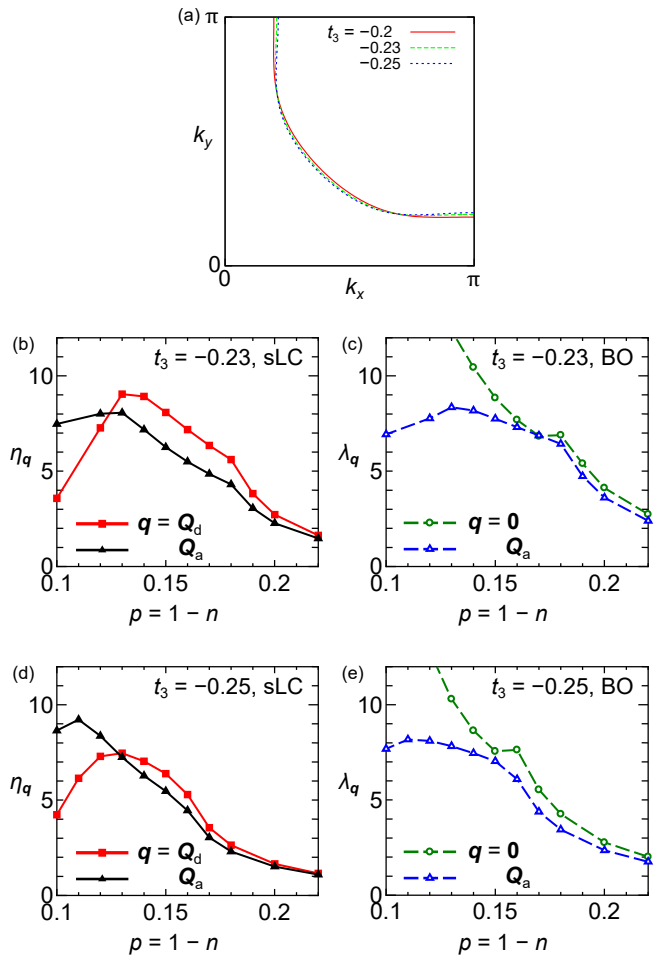}
\caption{
(a) FSs for $t_3=-0.20\sim-0.25$ at $n=0.85$, 
in the case of $t_1=-1.0$ and $t_2=1/6$.
%The shape of the flat part of the FS near the BZ 
%are sensitive to the parameter $t_3$.
The set $t_3=0.20$ in the main text.
(b)-(e) Obtained eigenvalue $\eta_\q$ and  $\lambda_\q$ 
in the cases of $t_3=-0.23$ ((b) and (c)) and 
$t_3=-0.25$ ((d) and (e)).
%(f) Schematic spin current pattern due to the sLC
%order at $\q=\Q_{\rm a}$.
}
\label{fig:phase-t3}
\end{figure}

Here, we examine a key model parameter for the phase diagram,
and reveal that the sLC instability is
sensitively controlled by $t_3$.
Figure \ref{fig:phase-t3} (a) shows the FSs for $t_3=-0.20\sim0.25$
in the case of $t_1=-1.0$ and $t_2=1/6$.
The shape of the flat part of the FS near the BZ,
which is significant for the density wave (DW) instabilities 
at $\q=\Q_{\rm a}, \Q_{\rm d}$, is sensitively modified by $t_3$.

%%%%%%%%%%%%%%%%%%%%
%%%%%%%%%%%%%%%%%%%

\begin{figure}[htb]
%\vspace{5mm}
\includegraphics[width=.99\linewidth]{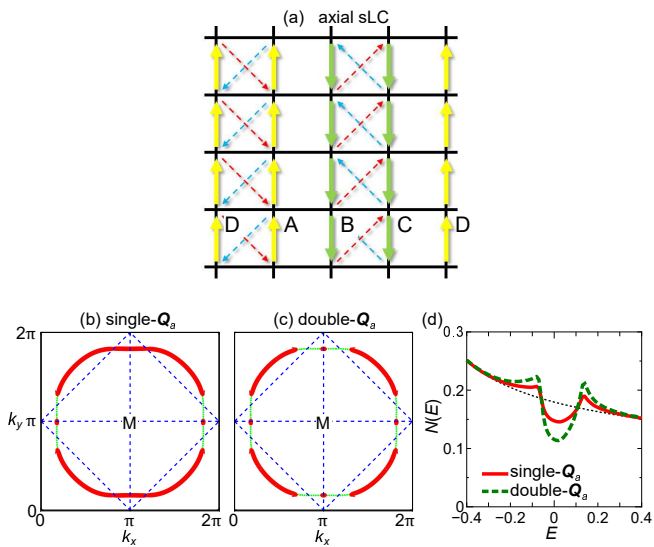}
\caption{
(a) Schematic spin current pattern due to the sLC
order at $\q=\Q_{\rm a}$.
(b) Fermi arc structure due to the single-$\q$ order,
(c) that due to the double-$\q$ order, and
(d) pseudogap in the DOS
due to the axial sLC order ($\q={\bm Q}_{\rm a}$).
%In both cases, we set $g^{\rm max}=0.1$.
}
\label{fig:axial-sLC}
\end{figure}

Figures \ref{fig:phase-t3} (b)-(e) shows the obtained
spin-channel eigenvalue $\eta_\q$ and 
the charge-channel one $\lambda_\q$ 
in the cases of $t_3=-0.23$ ((b) and (c)) and 
$t_3=-0.25$ ((d) and (e)).
(We set the condition $\a_S= 1-0.444p^2$ by following the main text.)
With increasing $|t_3|$, the peak of sLC instability is found to 
shift to the under-doped region.
Interestingly, the SLC eigenvalue at $\q=\Q_{\rm a}$ becomes 
larger than that at $\q=\Q_{\rm d}$ for $t_3=-0.25$.
Its spin current pattern in real space is shown in 
Fig. \ref{fig:axial-sLC} (a).

We note that recent experiments indicate that the phase diagram of 
cuprate superconductors is very diverse and rich. 
For example, the in-plain anisotropy of the ferro-magnetic 
susceptibility at $T=T^*$ is $B_{1g}$ in YBCO \cite{Y-Sato}, 
whereas $B_{2g}$ in Hg-compound \cite{Hg-Murayama}.  
Also, the antiferro-magnetic susceptibility in slightly 
under-doped YBCO exhibits a clear symmetry breaking at 
a temperate between $T^*$ and $T_{\rm CDW}$ \cite{Keimer}. 
The present sensitive $t_3$ dependent sLC
may give important hint to understand diverse symmetry breaking in cuprates.

%%%%%%%%%%%%%%%%%%%%
%%%%%%%%%%%%%%%%%%%

Here, we discuss the band-folding and hybridization gap
in the axial sLC phase.
Figures \ref{fig:axial-sLC} (b) and (c) show the Fermi arc structures
induced by the axial sLC order
in the cases of (a) single-$\q$ and (b) double-$\q$ orders.
%(The latter state is given by the coexistence of
%$\Q_{\rm a}=(\delta,0)$ and $\Q_{\rm a}'=(0,\delta)$ sLC order parameters.)
We set $g^{\rm max}\equiv \max_\k \{g_{\Q_{\rm d}}(\k)\} =0.1$.
Here, the folded band structure under the sLC order with 
finite $\q_{\rm sLC}$ is ``unfolded''
%into the original Brillouin zone by following Ref. \cite{Ku}
to make a comparison with ARPES results.
The Fermi arc due to the single-$\q$ order
in Fig. \ref{fig:axial-sLC} (b) belongs to the $B_{1g}$ symmetry.
%reflection symmetry about $k_x$- and $k_y$-axes.
%In contrast, the Fermi arc due to the double-$\q$ order
%in Fig. \ref{fig:fig5} (b) preserves the $C_4$ symmetry.
The resultant pseudogap in the DOS is shown in Fig. \ref{fig:axial-sLC} (d).
%The unfolded band structure in the single-$\Q_{\rm a}$ sLC order
%is displayed in Fig. \ref{fig:figS2} in the SM B \cite{SM}.
%ix \ref{sec:ApB}.
% \cite{SM}.
%The unfolded band structure is shown in Fig. S2 in the SM: B.

\subsection{Reduction of eigenvalues by $z<1$}

In the present work, we study the mechanism of exotic DW orders
due to the interference between paramagnons
based on the linearized DW equation.
The obtained form factor represents the characteristics and the symmetry of the DW,
and the eigenvalue expresses the strength of the DW instability.
In the present numerical study, we drop the self-energy in the DW equation. 
Then, the obtained eigenvalues shown in Fig. \ref{fig:fig4}
and Fig. \ref{fig:phase-t3} are much larger than unity.
In addition, the eigenvalue of the superconducting gap equation, $\l_{\rm SC}$,
shown in Fig. \ref{fig:fig4} (b) is also very large.

The self-energy gives the quasiparticle weight as
$z\equiv (1-{\rm Re}\d\Sigma(\e)/\d\e|_{\e=\mu})^{-1} \ (<1)$,
and $z^{-1}\ (>1)$ is the mass-enhancement.
The effect of the self-energy in the DW equation has been studied 
in Ref \cite{Onari-AFB} for Fe-based superconductors ($z\sim1/3$),
and it was revealed that the eigenvalue of the orbital fluctuations
is reduced by the self-energy, and the orbital order temperature
is reduced to $\sim100$K, consistently with experiments.

%%%%%%%%%%%%%%%%%%%%
\begin{figure}[htb]
%\vspace{5mm}
\includegraphics[width=.9\linewidth]{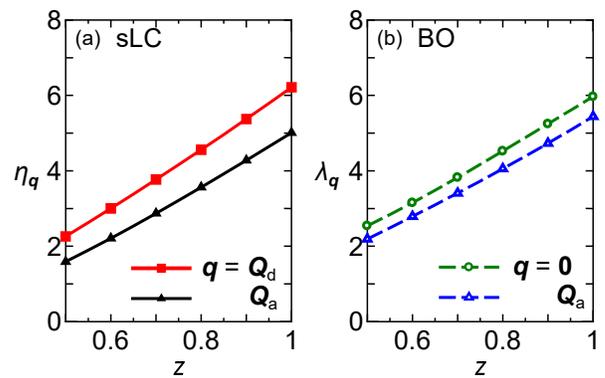}
\caption{
Obtained eigenvalues (a) $\eta_\q$ and (b) $\lambda_\q$ as functions of $z$.
}
\label{fig:eigen-z}
\end{figure}
%%%%%%%%%%%%%%%%%%%

Here, we study the effect of the renormalization factor $z$ 
on the eigenvalues in the present single-orbital Hubbard model.
Then, the Green function is given as
\begin{eqnarray}
G^z(k)=\frac1{i\e_n/z-\mu-\e_\k}.
\label{eqn:Gz}
\end{eqnarray}
First, we discuss the effect of $z$ on $\l_{\rm SC}$,
by replacing two $G$'s in Eq. (\ref{eqn:gap-eq}) with 
$G^z$ given by Eq. (\ref{eqn:Gz}).
On the other hand, we do not include $z$ in the susceptibilities 
$\chi^{s,c}$ in $V^{s,c}$ in the pairing interaction, 
in order not to change the Stoner factor $\a_S$.
Then, it is well known that $\l_{\rm SC}$ is reduced as $z\cdot \l_{\rm SC}$
based on the Eliashberg equation
\cite{Allen}.

Next, we discuss the effect of $z$ on $\lambda_\q$ ($\eta_\q$)
by replacing four $G$'s in Eqs. (\ref{eqn:DWeq1S}) and (\ref{eqn:IcS})
(in Eqs. (\ref{eqn:DWeq2S}) and (\ref{eqn:IsS}))
with $G^z$ given by Eq. (\ref{eqn:Gz}).
The obtained $z$-dependences of the eigenvalues at $T=0.03$ 
are shown in Fig. \ref{fig:eigen-z}.
It is verified that both $\lambda_\q$ and $\eta_\q$
are reduced in proportion to $z$.
Although this approximation may be justified only for $z\lesssim1$,
the obtained results strongly indicate that both $\lambda_\q$ and $\eta_\q$ 
are reduced to $O(1)$ in the case of $z\lesssim 0.2$,
which is realized in cuprate superconductors.

\subsection{Unfolded band structure under the sLC order}

%%%%%%%%%%%%%%%%%%%%
\begin{figure}[htb]
%\vspace{5mm}
\includegraphics[width=.99\linewidth]{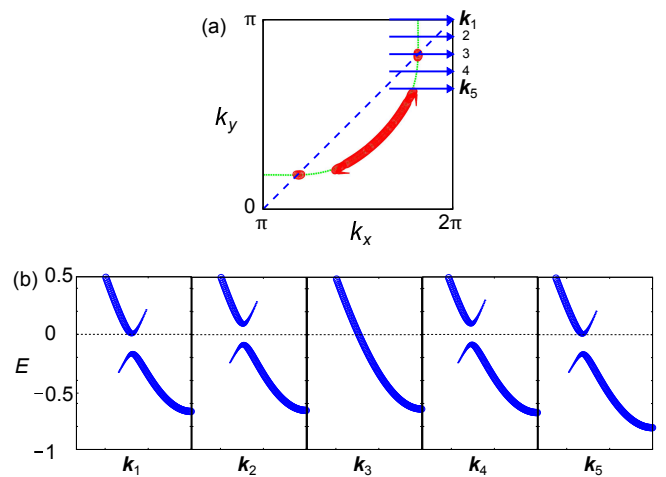}
\caption{
The unfolded band structure in the single-$\Q_{\rm d}$ sLC order
corresponds to Fig. \ref{fig:fig5} (a) in the main text.
}
\label{fig:figS2}
\end{figure}
%%%%%%%%%%%%%%%%%%%

Here, we examine the experimentally observed band structure
in the sLC ordered state, by applying the unfolding procedure
proposed in Ref. \cite{Ku}.
Figure \ref{fig:figS2}
shows the ``unfolded'' band structure in the single-$\Q_{\rm d}$ sLC order
at $g^{\rm max}=0.1$,
which corresponds to Fig. \ref{fig:fig5} (a) in the main text.
The pseudogap closes on the X-Y line owing to the odd-parity form factor.
This Dirac point which will be smeared out
for $T\sim T^* \ (\gg T_{\rm CDW})$
because of very large inelastic scattering at the hot spot
\cite{Kontani-ROP,Tremblay,Scalapino,Moriya}.
In addition, the Dirac point should be masked by
the $d$-wave BO below $T_{\rm CDW}$.

%%%%%%%%%%%%%%%%%%%%%%%%%%%%%%%%%%%%%%%%%%%%%%%%%%%%%%%%%%%%%%%
\section{BO/sLC order as magnon-pair condensation}
\label{sec:ApC}

%It is noteworthy that 
We explain that 
the sLC order is exactly the same as 
the magnon-pair condensation.
The following spin quadrupole order occurs
owing to the magnon-pair condensation
\cite{Andreev}:
%order parameter wiht time-reversal symmetry in real space:
%
\begin{eqnarray}
K_{i,j}^{\a,\b} &\equiv& \langle s_i^\a s_j^\b \rangle
-\langle s_i^\a s_j^\b \rangle_0,
\label{eqn:B1S}
%\\
%B_{i,j}^{\a,\b} &=& b_{i,j}^0 \delta_{\a,\b}
%+ \bm{b}_{i,j} \cdot (\hat{e}_\a \times \hat{e}_\b)
%%\bm{\e}_{\a,\b}
%\label{eqn:B2S}
\end{eqnarray}
where $\a,\b=x,y,z$, and 
the relation $K_{i,j}^{\a,\b}=K_{j,i}^{\b,\a}$ holds.
%$\bm{d}=(d^x,d^y,d^z)$, and
%$\hat{e}_\a$ is unit vector.
%$\bm{\e}_{\a,\b}=(\e_{x\a\b},\e_{y\a\b},\e_{z\a\b})$,
%where $\e_{\a\b\g}$ is the Levi-Civita tensor.
We will explain that the even-parity function
$a_{i,j} \equiv K_{i,j}^{\a,\a}/3$ \ (with $a_{i,j}=a_{j,i}$)
corresponds the BO state, and the odd-parity function
$b_{i,j}^\a\equiv i\e_{\a\b\g}K_{i,j}^{\b,\g}/2$ \
(with $b_{i,j}^\a=-b_{j,i}^\a$) corresponds the sLC order.

%%%%%%%%%%%%%%%%%%%%
\begin{figure}[htb]
%\vspace{5mm}
\includegraphics[width=.99\linewidth]{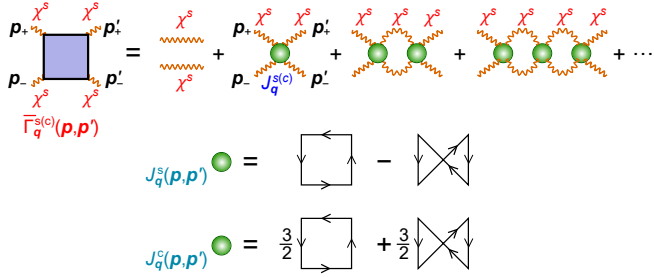}
\caption{
Diagrammatic expression of
$\bar{\Gamma}^{s(c)}_\q(p,p')$,
%G(k_+)G(k_-)\Gamma^{s(c)}_\q(k,k')G(k'_+)G(k'_-)$,
which represents the 
scattering process of the magnon pair
through the interaction $J_\q^{s(c)}(p,p')$.
Mathematically, $\bar{\Gamma}^{s(c)}_\q(p,p')$ diverges when 
magnon pairs with momentum $\q$ condense.
Thus, sLC/BO is interpreted as 
the condensation of odd/even parity magnon pairs.
}
\label{fig:figS3}
\end{figure}
%%%%%%%%%%%%%%%%%%

%$\Gamma^{s,c}_\q(k,k')$ due to AL processes
%represens the scattering process between two-magnons.
%For this purpose,

Here, we explain that 
$\Gamma^{s,c}_\q(k,k')$ due to the AL processes
represents the scattering between two-magnons.
To simplify the discussion,
we drop the MT term, and consider only AL terms with two $\chi^s$s.
Then, we define $\bar{\Gamma}^{s(c)}_\q(p,p')$ 
by the following relation;
$\Gamma^{c(s)}_\q(k,k')=T^2\sum_{p,p'}
[G(k-p)+(-)G(k+p)]\bar{\Gamma}^{c(s)}_\q(p,p')G(k'-p)$.
Figure \ref{fig:figS3} shows the
diagrammatic expression of $\bar{\Gamma}^{s,c}_\q(p,p')$,
%\equiv \chi^s(p_+)\chi^s(p_-)\Gamma^{s(c)}_\q(p,p')\chi^s(p'_+)\chi^s(p'_-)$,
which represents the 
scattering process of magnon pair amplitude $b^z$ ($a$)
through the interaction $J_\q^{s(c)}(p,p')$,
which is a moderate function of $T$.
With decreasing temperatures,
$\bar{\Gamma}^{c(s)}_\q(p,p')$ diverges when 
singlet (triplet) magnon pairs with momentum $\q$ condensate,
and the critical temperature corresponds to
$\lambda_\q=1$ ($\eta_\q=1$).

Here, we introduce
%$\bar{f}_\q(k)\equiv T\sum_p H_\q^+(k,p)f_\q(p)$ and
%$\bar{g}_\q(k)\equiv T\sum_p H_\q^-(k,p)g_\q(p)$,
%where $H_q^s(k,p)=G(p_+)G(p_-)[G(p-k)+s G(p+k)]$ ($s=\pm1$)
$\bar{f}_\q(k)\equiv T\sum_p H_\q(k,p)f_\q(p)$ and
$\bar{g}_\q(k)\equiv T\sum_p H_\q(k,p)g_\q(p)$,
where $H_q(k,p)=G(p_+)G(p_-)G(p-k)$,
and $f_\q(p)$ and $g_\q(p)$ are form factors of the DW equations.
Then, the DW equations are rewritten as
%Eqs. (\ref{eqn:DWeq1S}) and (\ref{eqn:DWeq2S}), 
% magnon pair condensations are
%
\begin{eqnarray}
\lambda_\q\bar{f}_\q(k)&=& T\sum_p J^c_\q(k,p)\chi^s(p_+)\chi^s(p_-)\bar{f}_\q(p)
\\
\eta_\q\bar{g}_\q(k)&=& T\sum_p J^s_\q(k,p)\chi^s(p_+)\chi^s(p_-)\bar{g}_\q(k)
\end{eqnarray}
where the kernel function $J^{c,s}_\q(k,p)$ is given in Fig. \ref{fig:figS3}.
These equations means that
$\bar{f}_\q(k)$ ($\bar{g}_\q(k)$) corresponds to 
the singlet (triplet) magnon pair condensation.
Therefore, their Fourier transformations correspond to
$a_{i,j}$ and $b_{i,j}^z$, respectively.

To summarize, in the present double spin-flip mechanism,
magnon-pair condensation $a,b^{z}\ne0$ occurs at $T=T_{\rm sLC}$.
Therefore, the sLC/BO given by the present mechanism 
is exactly the same as
``condensation of odd/even parity magnon pairs''.

%The order parameter $\bm{d}_\q(\k)$
%($d_\q^0(\k)$) is given by the Fourier transformation of 
%$\bm{d}_{i,j}$ ($d^0_{i,j}$) in Eq. (\ref{eqn:B2S}).

%%%%%%%%%%%%%%%%%%%%%%%%%%%%%%%%%%%%%%%%%%
\section{Spontaneous spin current in the sLC phase}
\label{sec:ApD}

\subsection{Calculation of the spin current}

Here, we investigate the spin current in real space
due to the ``spin-dependent self-energy'' $\delta t_{i,j}^\s = \s g_{i,j}$
shown in Fig. \ref{fig:fig3} (c), which is purely imaginary
and odd with respect to $i \leftrightarrow j$.
%Here, $g_{i,j}$ is the Fourier transformation of the form factor $g_{\q}(\k)$.
%Since $g_{\q}(\k)$ is odd-function, $\delta t_{i,j}^\s$ is purely imaginary.
%and the relation $\delta t_{i,j}^\uparrow=-\delta t_{j,i}^\downarrow$ holds.
%The conservation law $\dot{n}_i^\s = \sum_j j_{i,j}^\s$
%directly leads to the definition 
The spin current operator from site $j$ to site $i$ is
$j_{i,j}^\s =-i\sum_{\s}\s (h_{i,j}^\s c_{i\s}^\dagger c_{j\s}-(i \leftrightarrow j))$,
where $h_{i,j}^\s = t_{i,j}+\delta t_{i,j}^\s$.
Then, the spin current from $j$ to $i$ is given as
$J_{i,j}^s=\langle j_{i,j}^s \rangle_{\hat{h}^\s}$.

%%%%%%%%%%%%%%%%%%%%
\begin{figure}[htb]
%\vspace{5mm}
\includegraphics[width=.99\linewidth]{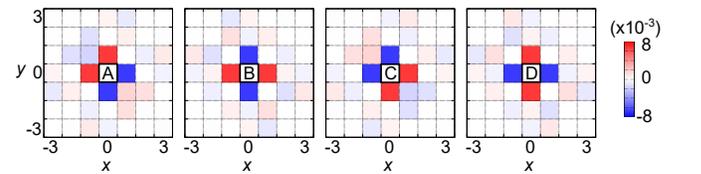}
\caption{
Obtained spin current from the center site (A-D) 
to different sites under the diagonal sLC with period $4 a_{\rm Cu-Cu}$.
The real space pattern is depicted in 
Fig. \ref{fig:fig1} (c), with sites A-D in a unit cell.
%(e)-(g) Fermi arc structures and pseudogap 
%due to the axial sLC order with $\q={\bm Q}_{\rm a}$.
%In calculating (a)-(d), we introduced BCS-type cutoff energy 
%$\w_c=0.5$ for the band-hybridization by $g_\q(\k)$.
}
\label{fig:S-current}
\end{figure}
%%%%%%%%%%%%%%%%%%%

Here, we calculate the spin current for the
commensurate sLC order at $\q_{\rm sLC}=(\pi/2,\pi/2)$,
which is achieved by putting $n=1.0$.
Then, the unit cell are composed four sites A-D.
Figure \ref{fig:S-current} shows the obtained spin current $J_{i,j}^s$
from the center site ($j={\rm A}$-${\rm B}$) to different site 
in Fig. \ref{fig:fig1} (c), by setting $g^{\rm max}=0.1$.
The obtained current is $|J_{i,j}^s|\sim10^{-2}$ 
in unit $|t_1|/\hbar$.
The derived spin current pattern 
between the nearest and second-nearest sites
is depicted in Fig. \ref{fig:fig1} (c).
The spin current is exactly conserved at each site.

%Previously, we showed one example of
%spin current pattern at $\q_{\rm sLC}=(\pi/2,\pi/2)$ 
%in Fig. \ref{fig:fig1} (c).

\subsection{Possible diagonal sLC patterns in real space}

%%%%%%%%%%%%%%%%%%%%
\begin{figure}[htb]
%\vspace{5mm}
\includegraphics[width=.7\linewidth]{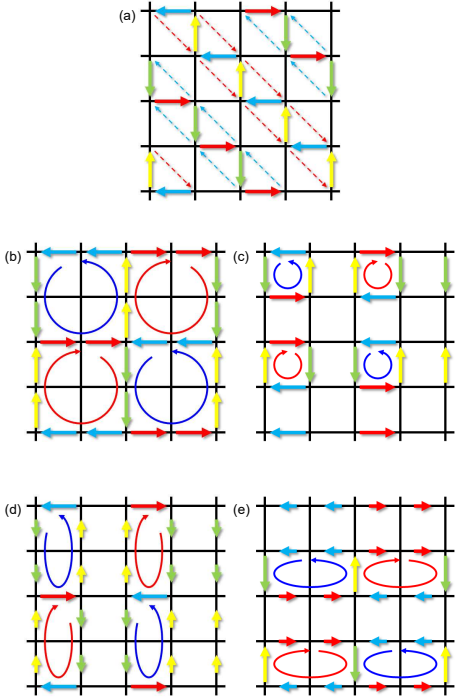}
\caption{
Examples of the diagonal sLC pattern in real space for 
$\q_{\rm sLC}=(\pi/2,\pi/2),(\pi/2,-\pi/2)$.
(a) Single-$\q$ sLC pattern for $\psi=\pi/2$.
(b)-(e) Four examples of double-$\q$ sLC patterns.
}
\label{fig:figS4}
\end{figure}
%%%%%%%%%%%%%%%%%%%

Next, we explain that 
the spin current pattern derived from the form factor 
$g_\q(\k)$ in Fig. \ref{fig:fig3} (b) is not uniquely determined.
In fact, the form factor in real space is given as
$i{\rm Im} \{g_{i,j}e^{i\psi}\} 
\sim i{\rm Im} \{e^{i\bm{q}\cdot (\bm{r}_i+\bm{r}_j)/2}e^{i\psi}\}$,
%${\rm Re} \sum_\k g_\q(\k) e^{i\k\cdot(\bm{r}_i-\bm{r}_j)} 
%e^{i\bm{q}\cdot (\bm{r}_i+\bm{r}_j)/2}e^{i\psi}$,
where $\psi$ is an arbitrary phase.
%${\rm Re}g_{i,j} e^{i\bm{u}\cdot (\bm{r}_i+\bm{r}_j)/2}e^{i\phi}$,
%where $f_{i,j}$ is the Fourier transformation of $g_\q(\k)$
%and $\bm{\phi}=(\phi_x,\phi_y)$ is arbitrary phase factor.
Here, we discuss other possible spin current patterns
by choosing $\psi$.

First, we discuss the real space pattern for 
$\q_{\rm sLC}=(\pi/2,\pi/2),(\pi/2,-\pi/2)$.
We assume that Fig. \ref{fig:fig1} (c) corresponds to $\psi=0$.
Then, the single-$\q$ spin current pattern for $\psi=\pi/2$
is given in Fig. \ref{fig:figS4} (a).
The double-$\q$ spin current order is given by the 
combination of the sLC order at 
$\q_{\rm sLC}=(\pi/2,\pi/2)$ and that at $\q_{\rm sLC}=(\pi/2,-\pi/2)$
with arbitrary phase factors.
Figure \ref{fig:figS4} (b)-(c) are given by the combination 
of Fig. 1 (c) with its $\pi/2$-rotation,
and Figs. \ref{fig:figS4} (d)-(e) are given by the combination 
of Figs. \ref{fig:figS4} (a) with its $\pi/2$-rotation.
We stress that
the magnitude of spin current $|J^s_{i,j}|$ 
in Fig. \ref{fig:figS4} (b)-(c) has $C_4$ symmetry,
whereas that in Fig. \ref{fig:figS4} (d)-(e) 
breaks the $C_4$ symmetry.% q=Qa,Qd
% different possibility: different phase (2 pictures)
% double q (4 pictures)

%%%%%%%%%%%%%%%%%%%%%%%%
%references
%%%%%%%%%%%%%%%%%%%%%%%%

%%%%%%%%%%%%%%%%%%%%%%%%%%%%%%%%%%%%%%%%%%%%%%%%%%


\begin{thebibliography}{99}


\bibitem{Y-Xray1}
G. Ghiringhelli, M. L. Tacon, M. Minola, S. Blanco-Canosa, C. Mazzoli, N. B. Brookes, G. M. D. Luca, A. Frano, D. G. Hawthorn, F. He, T. Loew, M. M. Sala, D. C. Peets, M. Salluzzo, E. Schierle, R. Sutarto, G. A. Sawatzky, E. Weschke, B. Keimer, and L. Braicovich, 
{\it Long-range incommensurate charge fluctuations in (Y,Nd)Ba$_2$Cu$_3$O$_{6+x}$},
Science {\bf 337}, 821 (2012).

\bibitem{Bi-Xray1}
%delta_c(n)
%delta_c = \Delta_{FS}
%d_{x2-y2}-CDW
R. Comin, A. Frano, M. M. Yee, Y. Yoshida, H. Eisaki, E. Schierle, E. Weschke, . Sutarto, F. He, A. Soumyanarayanan, Y. He, M. L. Tacon, I. S. Elfimov, J. E. Hffman, G. A. Sawatzky, B. Keimer, and A. Damascelli,
{\it Charge order driven by Fermi-arc instability in Bi$_2$Sr$_{2-x}$La$_x$CuO$_{+\delta}$},
Science {\bf 343}, 390 (2014).

\bibitem{STM-Kohsaka}
Y. Kohsaka, T. Hanaguri, M. Azuma, M. Takano, J. C. Davis, and H. Takagi,
{\it Visualization of the emergence of the pseudogap state and the evolution to superconductivity in a lightly hole-doped Mott insulator},
Nature Physics {\bf 8}, 534 (2012). 
%Visualization of the emergence of the pseudogap state and the evolution to superconductivity in a lightly hole-doped Mott insulator 

\bibitem{STM-Fujita}
K. Fujita, M. H. Hamidian, S. D. Edkins, C. K. Kim, Y. Kohsaka, M. Azuma, M. Takano, H. Takagi, H. Eisaki, S. Uchida, A. Allais, M. J. Lawler, E.-A. Kim, S. Sachdev, and J. C. S. Davis, 
{\it Direct phase-sensitive identification of a $d$-form factor density wave in underdoped cuprates},
Proc. Natl. Acad. Sci. USA, {\bf 111}, E3026 (2014).
%arXiv:1404.0362.

\bibitem{Tremblay}
B. Kyung, S. S. Kancharla, D. Senechal, A.-M. S. Tremblay, M. Civelli, and G. Kotliar, 
{\it Pseudogap induced by short-range spin correlations in a doped Mott insulator},
Phys. Rev. B {\bf 73}, 165114 (2006).

\bibitem{Scalapino}
T.A. Maier, M.S. Jarrell, and D.J. Scalapino, 
{\it Understanding high-temperature superconductors with quantum cluster theories},
Physica C, {\bf 460-462}, 13 (2007).

\bibitem{Moriya}
T. Moriya and K. Ueda, 
{\it Spin fluctuations and high temperature superconductivity},
Adv. Phys. {\bf 49}, 555 (2000).

\bibitem{RUS}
A.Shekhter, B. J. Ramshaw, R. Liang, W. N. Hardy, D. A. Bonn,
F. F. Balakirev, R. D. McDonald, J. B. Betts, S. C. Riggs, and A. Migliori,
{\it Bounding the pseudogap with a line of phase transitions in YBa$_2$Cu$_3$O$_{6+\delta}$},
Nature {\bf 498}, 75 (2013).

\bibitem{ARPES-Science2011}
R.-H. He, M. Hashimoto, H. Karapetyan, J. D. Koralek, J. P. Hinton, J. P. Testaud, V. Nathan, Y. Yoshida, H. Yao, K. Tanaka, W. Meevasana, R. G. Moore, D. H. Lu, S.-K. Mo, M. Ishikado, H. Eisaki, Z. Hussain, T. P. Devereaux, S. A. Kivelson, J. Orenstein, A. Kapitulnik, and Z.-X. Shen,
{\it From a Single-Band Metal to a High-Temperature Superconductor via Two Thermal Phase Transitions},
Science {\bf 331}, 1579 (2011).

\bibitem{Y-Sato}
Y. Sato, S. Kasahara, H. Murayama, Y. Kasahara, E. -G. Moon, T. Nishizaki, T. Loew, J. Porras, B. Keimer, T. Shibauchi, and Y. Matsuda,
{\it Thermodynamic evidence for a nematic phase transition at the onset of the pseudogap in YBa$_2$Cu$_3$O$_y$},
Nature Physics {\bf 13}, 1074 (2017).

\bibitem{Hg-Murayama}
H. Murayama, Y. Sato, R. Kurihara, S. Kasahara, Y. Mizukami, Y. Kasahara, H. Uchiyama, A. Yamamoto, E. -G. Moon, J. Cai, J. Freyermuth, M. Greven, T. Shibauchi, and Y. Matsuda,
{\it Diagonal nematicity in the pseudogap phase of HgBa$_2$CuO$_{4+\delta}$},
Nat. Commun. {\bf 10}, 3282 (2019).

\bibitem{Fujimori-nematic}
S. Nakata, M. Horio, K. Koshiishi, K. Hagiwara, C. Lin, M. Suzuki, S. Ideta, K. Tanaka, D. Song, Y. Yoshida, H. Eisaki, and A. Fujimori,
{\it Nematicity in the pseudogap state of cuprate superconductors revealed by angle-resolved photoemission spectroscopy},
arXiv:1811.10028.

\bibitem{Shibauchi-nematic}
K. Ishida, S. Hosoi, Y. Teramoto, T. Usui, Y. Mizukami, K. Itaka, Y. Matsuda, T. Watanabe, and T. Shibauchi,
{\it Divergent nematic susceptibility near the pseudogap critical point in a cuprate superconductor},
J. Phys. Soc. Jpn. {\bf 89}, 064707 (2020).

\bibitem{Bulut}
S. Bulut, W.A. Atkinson and A.P. Kampf,
{\it Spatially modulated electronic nematicity in the three-band model of cuprate superconductors}
Phys. Rev. B {\bf 88}, 155132 (2013).

\bibitem{Chubukov}
%spin-fermion model
Y. Wang and A.V. Chubukov, 
{\it Charge-density-wave order with momentum $(2Q,0)$ and $(0,2Q)$ within the spin-fermion model: Continuous and discrete symmetry breaking, preemptive composite order, and relation to pseudogap in hole-doped cuprates},
Phys. Rev. B {\bf 90}, 035149 (2014).

\bibitem{Chubukov-AL}
R.-Q. Xing, L. Classen, and A. V. Chubukov,
{\it Orbital order in FeSe: The case for vertex renormalization},
Phys. Rev. B {\bf 98}, 041108(R) (2018).

\bibitem{Sachdev}
%spin-fermion model
M.A. Metlitski and S. Sachdev, 
{\it Instabilities near the onset of spin density wave order in metals},
New J. Phys. {\bf 12}, 105007 (2010);
%t-J; MF
S. Sachdev and R. La Placa, 
{\it Bond Order in Two-Dimensional Metals with Antiferromagnetic Exchange Interactions},
Phys. Rev. Lett. {\bf 111}, 027202 (2013).
%S. Whitsitt and S. Sachdev, arXiv:1406.6061.

\bibitem{Metzner}
W. Metzner, M. Salmhofer, C. Honerkamp, V. Meden, and K. Sch\"{o}nhammer, 
{\it Functional renormalization group approach to correlated fermion systems},
Rev. Mod. Phys. {\bf 84}, 299 (2012);
C. Husemann and W. Metzner, 
{\it Incommensurate nematic fluctuations in the two-dimensional Hubbard model},
Phys. Rev. B {\bf 86}, 085113 (2012);
T. Holder and W. Metzner, 
{\it Incommensurate nematic fluctuations in two-dimensional metals},
Phys. Rev. B {\bf 85}, 165130 (2012);
C. Honerkamp, 
{\it Charge instabilities at the metamagnetic transition of itinerant electron systems},
Phys. Rev. B {\bf 72}, 115103 (2005).

\bibitem{DHLee-PNAS}
%t-J
J. C. S. Davis and D.-H. Lee, 
{\it Concepts relating magnetic interactions, intertwined electronic orders, and strongly correlated superconductivity},
Proc. Natl. Acad. Sci. USA, {\bf 110}, 17623 (2013).

\bibitem{Kivelson-NJP}
%t-J
E. Berg, E. Fradkin, S. A. Kivelson, and J. M. Tranquada, 
{\it Striped superconductors: how spin, charge and superconducting orders intertwine in the cuprates},
New J. Phys. {\bf 11}, 115004 (2009).

\bibitem{Yamakawa-CDW}
Y. Yamakawa, and H. Kontani,
{\it Spin-Fluctuation-Driven Nematic Charge-Density Wave in Cuprate Superconductors: Impact of Aslamazov-Larkin Vertex Corrections},
Phys. Rev. Lett. {\bf 114}, 257001 (2015).

\bibitem{Tsuchiizu-CDW}
M. Tsuchiizu, K. Kawaguchi, Y. Yamakawa, and H. Kontani,
{\it Multistage electronic nematic transitions in cuprate superconductors: A functional-renormalization-group analysis},
Phys. Rev. B {\bf 97}, 165131 (2018).
%Phys. Rev. B {\bf 93}, 155148 (2016).

\bibitem{Kawaguchi-CDW}
K. Kawaguchi, Y. Yamakawa, M. Tsuchiizu, and H. Kontani,
{\it Competing Unconventional Charge-Density-Wave States in Cuprate Superconductors: Spin-Fluctuation-Driven Mechanism},
J. Phys. Soc. Jpn. {\bf 86}, 063707 (2017).

\bibitem{PALee}
P. A. Lee,
{\it Amperean Pairing and the Pseudogap Phase of Cuprate Superconductors},
Phys. Rev. X {\bf 4}, 031017 (2014).

\bibitem{Agterberg}
D. F. Agterberg, J. C. S. Davis, S. D. Edkins, E. Fradkin, D. J. Van Harlingen, S. A. Kivelson, P. A. Lee, L. Radzihovsky, J. M. Tranquada, and Y. Wang,
{\it The Physics of Pair Density Waves},
Annu. Rev. Condens. Matter Phys. {\bf 11}, 231 (2020).

\bibitem{Varma}
C. M. Varma, 
{\it Non-Fermi-liquid states and pairing instability of a general model of copper oxide metals},
Phys. Rev. B {\bf 55}, 14554 (1997).
%{\it Theory of the pseudogap state of the cuprates},
%Phys. Rev. B {\bf 73}, 155113 (2006).

\bibitem{Affleck}
I. Affleck and J. B. Marston, 
{\it Large-$n$ limit of the Heisenberg-Hubbard model: Implications for high-$T_c$ superconductors},
Phys. Rev. B {\bf 37}, 3774(R) (1988).

\bibitem{FCZhang}
F. C. Zhang, 
{\it Superconducting instability of staggered-flux phase in the t-J model},
Phys. Rev. Lett. {\bf 64}, 974 (1990).

\bibitem{Schultz}
H. J. Schulz, 
{\it Fermi-surface instabilities of a generalized two-dimensional Hubbard model},
Phys. Rev. B {\bf 39}, 2940(R) (1989).

\bibitem{Nersesyan}
A. A. Nersesyan, G. I. Japaridze, and I. G. Kimeridze, 
{\it Low-temperature magnetic properties of a two-dimensional spin nematic state},
J. Phys.: Condens. Matter {\bf 3}, 3353 (1991).

\bibitem{Ozaki}
M. Ozaki, 
{\it Broken Symmetry Solutions of the Extended Hubbard Model},
Int. J. Quantum. Chem. {\bf 42}, 55 (1992).

\bibitem{Sr2IrO4}
S. Zhou, K. Jiang, H. Chen, and Z. Wang,
{\it Correlation Effects and Hidden Spin-Orbit Entangled Electronic Order in Parent and Electron-Doped Iridates Sr$_2$IrO$_4$},
Phys. Rev. X {\bf 7}, 041018 (2017).


\bibitem{Ikeda}
H. Ikeda and Y. Ohashi, 
{\it Theory of Unconventional Spin Density Wave: A Possible Mechanism of the Micromagnetism in U-based Heavy Fermion Compounds},
Phys. Rev. Lett. {\bf 81}, 3723 (1998).

\bibitem{Fujimoto}
S. Fujimoto, 
{\it Spin Nematic State as a Candidate of the Hidden Order Phase of URu$_2$Si$_2$},
Phys. Rev. Lett. {\bf 106}, 196407 (2011).

\bibitem{Onari-SCVC}
S. Onari and H. Kontani, 
{\it Self-consistent Vertex Correction Analysis for Iron-based Superconductors: Mechanism of Coulomb Interaction-Driven Orbital Fluctuations},
Phys. Rev. Lett. {\bf 109}, 137001 (2012).

\bibitem{Onari-FeSe}
S. Onari, Y. Yamakawa, and H. Kontani,
{\it Sign-Reversing Orbital Polarization in the Nematic Phase of FeSe due to the $C_2$ Symmetry Breaking in the Self-Energy},
Phys. Rev. Lett. {\bf 116}, 227001 (2016).

\bibitem{Yamakawa-FeSe}
Y. Yamakawa, S. Onari, and H. Kontani,
{\it Nematicity and Magnetism in FeSe and Other Families of Fe-Based Superconductors},
Phys. Rev. X {\bf 6}, 021032 (2016).

\bibitem{Yoshida-arc}
T. Yoshida, M. Hashimoto, I. M. Vishik, Z. X. Shen, and A. Fujimori,
{\it Pseudogap, Superconducting Gap, and Fermi Arc in High-$T_c$ Cuprates Revealed by Angle-Resolved Photoemission Spectroscopy},
J. Phys. Soc. Jpn. {\bf 81}, 011006 (2011).

\bibitem{Sr2IrO4-ARPES}
Y. K. Kim, O. Krupin, J. D. Denlinger, A. Bostwick, E. Rotenberg, Q. Zhao, J. F. Mitchell, J. W. Allen, an B. J. Kim,
{\it Fermi arcs in a doped pseudospin-1/2 Heisenberg antiferromagnet},
Science {\bf 345}, 187 (2014).

\bibitem{Kontani-ROP}
H. Kontani, 
{\it Anomalous Transport Phenomena in Fermi Liquids with Strong Magnetic Fluctuations},
Rep. Prog. Phys. {\bf 71}, 026501 (2008);

\bibitem{Springer}
H. Kontani, {\it Transport Phenomena in Strongly Correlated Fermi Liquids}
(Springer-Verlag Berlin and Heidelberg GmbH \& Co. K, 2013).

\bibitem{neutron}
C. Stock, W. J. L. Buyers, R. Liang, D. Peets, Z. Tun, D. Bonn, 
W. N. Hardy, and R. J. Birgeneau,
{\it Dynamic stripes and resonance in the superconducting and normal phases of YBa$_2$Cu$_3$O$_{6.5}$ ortho-II superconductor},
Phys. Rev. B {\bf 69}, 014502 (2004).

%aaa
\bibitem{Tazai-CeCu2Si2}
{\it Hexadecapole Fluctuation Mechanism for s-wave Heavy Fermion Superconductor CeCu$_2$Si$_2$: Interplay between Intra- and Inter-Orbital Cooper Pairs}
J. Phys. Soc. Jpn. {\bf 88}, 063701 (2019);
R. Tazai, Y. Yamakawa, and H. Kontani,
{\it Plain s-wave superconductivity near the magnetic criticality: Enhancement of attractive electron-boson coupling vertex corrections},
J. Phys. Soc. Jpn. {\bf 86}, 073703 (2017).

\bibitem{Tazai-CeB6}
R. Tazai and H. Kontani,
{\it Multipole fluctuation theory for heavy fermion systems: Application to multipole orders in CeB$_6$},
Phys. Rev. B {\bf 100}, 241103(R) (2019).

\bibitem{Tsuchiizu-PRL}
M. Tsuchiizu, Y. Ohno, S. Onari, and H. Kontani,
{\it Orbital Nematic Instability in the Two-Orbital Hubbard Model: Renormalization-Group + Constrained RPA Analysis}, 
Phys. Rev. Lett. {\bf 111}, 057003 (2013).

\bibitem{Tazai-FRG}
R. Tazai, Y. Yamakawa, M. Tsuchiizu, and  H. Kontani,
{\it Functional renormalization group study of orbital fluctuation mediated superconductivity: Impact of the electron-boson coupling vertex corrections},
Phys. Rev. B {\bf 94}, 115155 (2016).

\bibitem{Tazai-kappa}
R. Tazai, Y. Yamakawa, M. Tsuchiizu, and  H. Kontani,
{\it Prediction of $d$-wave bond-order and pseudogap in organic superconductor $\kappa$-(BEDT-TTF)$_2$X: Similarities to cuprate superconductors},
arXiv:2010.15516.

\bibitem{Tazai-cLC}
R. Tazai, Y. Yamakawa, and  H. Kontani,
{\it Emergence of Charge Loop Current in Geometrically Frustrated Hubbard Model: Functional Renormalization Group Study},
arXiv:2010.16109.

\bibitem{Onari-B2g}
S. Onari and H. Kontani,
{\it Origin of diverse nematic orders in Fe-based superconductors: 45 degree rotated nematicity in AFe$_2$As$_2$ (A=Sc,Rb)},
Phys. Rev. B {\bf 100}, 020507(R) (2019).

\bibitem{Onari-AFB}
S. Onari and H. Kontani,
{\it Hidden antiferro-nematic order in Fe-based superconductor BaFe$_2$As$_2$ and NaFeAs above $T_S$},
Phys. Rev. Research {\bf 2}, 042005 (2020).

\bibitem{d-wave}
D. J. Scalapino, E. Loh, and J. E. Hirsch, 
{\it d-wave pairing near a spin-density-wave instability}, 
Phys. Rev. B {\bf 34}, 8190 (1986).

\bibitem{Norman}
V. Mishra and M. R. Norman, 
{\it Strong coupling critique of spin fluctuation driven charge order in underdoped cuprates},
Phys. Rev. B {\bf 92}, 060507(R) (2015).

\bibitem{Andreev}
A. F. Andreev and I. A. Grishchuk, 
{\it Spin nematics},
Zh. Eksp. Teor. Fiz. {\bf 87}, 467 (1984).

\bibitem{Coleman}
P. Chandra and P. Coleman,
{\it Quantum spin nematics: Moment-free magnetism},
Phys. Rev. Lett. {\bf 66}, 100 (1991).

\bibitem{Shannon}
N. Shannon, T. Momoi, and P. Sindzingre,
{\it Nematic Order in Square Lattice Frustrated Ferromagnets},
Phys. Rev. Lett. {\bf 96}, 027213 (2006).

\bibitem{Ku}
W. Ku, T. Berlijn, and C.-C. Lee,
{\it Unfolding First-Principles Band Structures},
Phys. Rev. Lett. {\bf 104}, 216401 (2010).

\bibitem{Kawaguchi-B2g}
K. Kawaguchi and H. Kontani, 
{\it Spin-fluctuation-driven $B_{1g}$ and $B_{2g}$ Bond Order and Induced In-plane Anisotropy in Magnetic Susceptibility in Cuprate Superconductors: Impact of Hot-/Cold-spot Structure on Bond-order Symmetry},
J. Phys. Soc. Jpn. {\bf 89}, 124704 (2020).

\bibitem{Keimer} 
V. Hinkov, D. Haug, B. Fauque, P. Bourges, Y. Sidis, A. Ivanov, C. Bernhard, C. T. Lin, and B. Keimer,
{\it Electronic Liquid Crystal State in the High-Temperature Superconductor YBa$_2$Cu$_3$O$_{6.45}$},
Science {\bf 319}, 597 (2008).

\bibitem{Allen} 
P. B. Allen and B. Mitrovic,
{\it Theory of Superconducting $T_c$},
Solid State Physics, {\bf 37}, 1  (1983)


\end{thebibliography}
\end{document}